\documentclass[aps,pre,reprint,superscriptaddress,fleqn,floatfix]{revtex4-2}
\usepackage{graphicx}
\usepackage{dcolumn}
\usepackage{bm}
\usepackage{amsmath, amsthm, amsxtra, amsfonts, amssymb,amscd,mathtools}
\usepackage{latexsym}
\usepackage{color}
\usepackage{enumitem}

\usepackage{etoolbox}
\usepackage{autobreak}
\usepackage[normalem]{ulem}
\usepackage{algorithm}
\usepackage{algpseudocode}
\usepackage{xr}

\def\del{{\partial}}
\def\R{{\mathbb R}}
\DeclareMathOperator*{\argmax}{arg\,max} 
\DeclareMathOperator*{\argmin}{arg\,min}

\usepackage[
  hidelinks, 
  ]{hyperref}

\begin{document}


\title{Bayesian estimation of coupling strength and heterogeneity in a coupled oscillator model from macroscopic quantities}


\author{Yusuke Kato}
\email{yuukato@g.ecc.u-tokyo.ac.jp}
\affiliation{Department of Complexity Science and Engineering, Graduate School of Frontier Sciences, The University of Tokyo, 5-1-5 Kashiwanoha, Kashiwa, Chiba 277-8561, Japan}

\author{Shuhei Kashiwamura}
\affiliation{Department of Physics, Graduate School of Science, The University of Tokyo, 7-3-1 Hongo, Bunkyo-ku, Tokyo 113-0033, Japan}

\author{Emiri Watanabe}
\affiliation{Department of Biological Sciences, Graduate School of Science, The University of Tokyo, 7-3-1 Hongo, Bunkyo-ku, Tokyo 113-0033, Japan}

\author{Masato Okada}
\affiliation{Department of Complexity Science and Engineering, Graduate School of Frontier Sciences, The University of Tokyo, 5-1-5 Kashiwanoha, Kashiwa, Chiba 277-8561, Japan}
\affiliation{Department of Physics, Graduate School of Science, The University of Tokyo, 7-3-1 Hongo, Bunkyo-ku, Tokyo 113-0033, Japan}

\author{Hiroshi Kori}
\affiliation{Department of Complexity Science and Engineering, Graduate School of Frontier Sciences, The University of Tokyo, 5-1-5 Kashiwanoha, Kashiwa, Chiba 277-8561, Japan}


\date{\today}

\begin{abstract}
Various macroscopic oscillations, such as the heartbeat and the flushing of fireflies, are created by synchronizing oscillatory units (oscillators). To elucidate the mechanism of synchronization, several coupled oscillator models have been devised and extensively analyzed. Although parameter estimation of these models has also been actively investigated, most of the proposed methods are based on the data from individual oscillators, not from macroscopic quantities. In the present study, we propose a Bayesian framework to estimate the model parameters of coupled oscillator models, using the time series data of the Kuramoto order parameter as the only given data. We adopt the exchange Monte Carlo method for the efficient estimation of the posterior distribution and marginal likelihood. Numerical experiments are performed to confirm the validity of our method and examine the dependence of the estimation error on the observational noise and system size. 
\end{abstract}


\maketitle

\section{Introduction}
Just as the collective motion of the heart originates from the cooperation of individual cardiomyocytes, some oscillations in nature and society emerge from the synchronization of multiple interacting units (oscillators) \cite{pikovsky2002synchronization}. Examples include the rhythmic flushing of fireflies \cite{smith1935synchronous}, pathologic synchronized brain activities in several neurological disorders \cite{uhlhaas2006neural, hammond2007pathological, jiruska2013synchronization}, and the spontaneous large-scale vibration of London's Millennium Bridge caused by coordinated pedestrian movement \cite{strogatz2005crowd}. 

To elucidate the mechanism of synchronization, various coupled phase-oscillator models, such as the Kuramoto model \cite{kuramoto84}, have been devised and extensively analyzed in a theoretical context \cite{acebron2005kuramoto, rodrigues2016kuramoto, pietras2019network}.  In addition to theoretical interest, such models have been applied to practical problems, including understanding neuronal disorders \cite{cumin2007generalising}, analyzing power grid dynamics \cite{filatrella2008analysis, potratzki2024synchronization}, and designing high-speed computers (coherent Ising machines) \cite{wang2017oscillator}. 

Among the parameters in coupled oscillator models, two are particularly crucial for achieving synchrony: the strength of the interaction between the oscillators (coupling strength) and the variation among the oscillator population (heterogeneity). If the heterogeneity is large relative to the coupling strength, synchronization is less likely to occur, and vice versa. Estimating these parameters from observed data is therefore essential to understand and control real-world synchronization phenomena.

Although many previous studies have explored data-driven parameter estimation in coupled oscillator systems \cite{smelyanskiy2005reconstruction, stankovski2015coupling, stankovski2017coupling, tokuda2019practical, matsuki2024network, su2025pairwise}, most rely on time series data from all individual oscillators. In practical settings, however, the number of observable oscillating units (e.g., cells) is often limited. Furthermore, the computational cost of estimation increases with the size of the dataset. These limitations motivate the development of estimation methods based on macroscopic quantities that capture collective oscillatory behavior, rather than requiring detailed observations of individual oscillators.

One such macroscopic quantity is the Kuramoto order parameter, which is widely used to quantify the degree of synchronization \cite{pikovsky2002synchronization, rodrigues2016kuramoto, ozawa2024two}. 
Within a theoretical framework called the Ott-Antonsen ansatz \cite{ott2008low, ott2009long}, the dynamics of the Kuramoto order parameter can be analytically derived for several coupled oscillator models in the thermodynamic limit.
By using this analytical solution, we expect that efficient parameter estimation can be achieved. Although a recent study \cite{yamaguchi2024reconstruction} employed the linear and nonlinear response theory for the parameter estimation from the time series of the generalized order parameter (Daido order parameter \cite{daido1992order}), no previous study, to the best of our knowledge, has attempted to estimate model parameters by directly using the analytical results from the Ott-Antonsen ansatz.

Among various estimation methods, Bayesian inference offers a key advantage in providing both the posterior distribution of parameters and the marginal likelihood. The posterior distribution allows for the construction of confidence intervals for estimated parameters, and the marginal likelihood quantifies the plausibility of the model for a given dataset, enabling model comparison and selection. We expect that adopting a Bayesian approach leads to accurate and reliable model estimations.

In the present study, we aim to use Bayesian inference to estimate the model parameters of the Kuramoto model using the time series data of the Kuramoto order parameter. The analytical solution derived via the Ott–Antonsen ansatz serves as the forward model in the estimation process.

To enhance estimation accuracy and efficiency, we employ the exchange Monte Carlo (EMC) method to compute the posterior distribution and marginal likelihood. Originally introduced in the field of statistical physics \cite{hukushima1996exchange}, the EMC method has since been applied to a wide range of Bayesian inference problems \cite{nagata2012bayesian, tokuda2017simultaneous, kashiwamura2022bayesian, ueda2023bayesian, kashiwamura2024noise}. In the context of coupled oscillator systems, although a previous study applied the EMC method to network optimization \cite{yanagita2010design}, its use for parameter estimation has not, to the best of our knowledge, been explored. This study thus represents the first attempt to estimate the parameters of coupled oscillator models by combining the theoretical results from the Ott–Antonsen ansatz with a Bayesian framework based on the EMC method.

The remainder of this paper is organized as follows: in Sec.~\ref{sec:pre}, we describe the model and the estimation framework. We introduce the Kuramoto model and its analytical solution based on the Ott-Antonsen ansatz in Sec.~\ref{sec:model}, and then we describe a detailed explanation of the EMC algorithm in Sec.~\ref{sec:frame}. The results of numerical experiments are presented in Sec.~\ref{sec:result}, in which we generate artificial datasets and examine how the estimation accuracy depends on the observational noise and system size. 
We first consider a benchmark case where the models for data generation and estimation are identical, to validate the consistency of our framework (Sec.~\ref{sec:res_oa}). Next, we analyze the case where datasets are generated from the Kuramoto model \eqref{eq:kura} and evaluate the estimation performance for various system sizes (Sec.~\ref{sec:res_kura}). Finally, in Sec.~\ref{sec:dis}, we discuss the validity of the estimation model [Eq.~ \eqref{eq:data}] and propose a modified formulation [Eq.~ \eqref{eq:data2}] to improve parameter inference for the Kuramoto model.

\section{Theory}
\label{sec:pre}
\subsection{Model}
\label{sec:model}
\subsubsection{Kuramoto model and Kuramoto order parameter}
We consider the following Kuramoto model: 
\begin{equation}
    \label{eq:kura}
    \dot{\phi_i} \coloneqq \frac{d \phi_i}{dt} = \omega_i + \frac{K}{N}\sum_{j=1}^{N}\sin (\phi_j - \phi_i), 
\end{equation}
for $i = 1, 2, \ldots, N$. Here, $\phi_i(t)$ and $\omega_i$ denote the phase and the natural frequency of the $i$-th oscillator, $K$ the strength of the coupling among oscillators, and $N$ the number of oscillators, respectively. We assume that $\omega_i$ is randomly drawn from a given probability density function $g(\omega)$. 

The following real variables $R$ and $\Phi$ are  commonly used to investigate the collective behavior of the Kuramoto model~\eqref{eq:kura}:
\begin{equation}
    R(t) e^{i \Phi(t)} \coloneqq \frac{1}{N}\sum_{k=1}^{N}e^{i \phi_k (t)}. 
\end{equation}
Here, the variable $R$, which satisfies $0\leq R \leq 1$, represents the degree of synchronization of the whole oscillators and is called the Kuramoto order parameter~\cite{rodrigues2016kuramoto, ozawa2024two}. If $R=1$ each oscillator has the same phase, meaning that a complete synchrony is achieved, whereas if $R=0$ their phases are uniformly distributed and not synchronized. The other variable $\Phi$ denotes the mean phase of the oscillator population. 

In what follows, we call the Kuramoto order parameter $R(t)$ as the order parameter for simplicity.

\subsubsection{Analytical solution by Ott-Antonsen Ansatz}
Ott and Antonsen found that, in the thermodynamic limit of $N \to \infty$, the long-time behavior of Eq.~\eqref{eq:kura} is confined to an invariant manifold (the Ott-Antonsen manifold) under certain assumptions on $g(\omega)$ \cite{ott2008low, ott2009long}. In addition, if $g(\omega)$ is Lorentzian, they found that the time-evolution of the order parameter $R$ on this manifold can be derived in a closed form \cite{ott2008low}. 

In the present study, we assume the following Lorentzian as the distribution of natural frequencies: 
\begin{equation}
    \label{eq:lorentz}
    g(\omega) = \frac{\gamma}{\pi(\omega^2 + \gamma^2)},
\end{equation}
where $\gamma$ is a positive constant. 
Then, according to the theory by Ott and Antonsen \cite{ott2008low}, the dynamics of order parameter $R(t)$ in Eq.~\eqref{eq:kura} is given as
\begin{equation}
    \label{eq:oa}
    \dot R = -\lambda R + (\lambda - \gamma) R^3, 
\end{equation}
where 
\begin{equation}
    \label{eq:lam}
    \lambda \coloneqq \gamma - \frac{K}{2}.
\end{equation}
The non-constant solution of Eq.~\eqref{eq:oa} is
\begin{equation}
    \label{eq:oa_sol_0}
    R(t) = R_{\rm sol}(t; \theta) \coloneqq \frac{e^{-\lambda t}}{\sqrt{\frac{1}{R_0^2} + \frac{\lambda - \gamma}{\lambda}\left( e^{-2 \lambda t} - 1 \right)}},
\end{equation}
with $R_0 \coloneqq R(0)$. Here, $\theta$ denotes the set of model parameters, i.e., 
\begin{equation}
\label{eq:theta}
    \theta \coloneqq \{ \gamma, \lambda, R_0 \}.
\end{equation}
Note that we treat the initial condition $R_0$ as one of the model parameters. 

\subsection{Framework for Bayesian estimation}
\label{sec:frame}
In this section, we explain the algorithm to estimate the set of model parameters $\theta$ from the time-series data of the order parameter $R(t)$. When describing the formulation of the Bayesian estimation and the EMC method, we consult Refs.~\cite{nagata2012bayesian, tokuda2017simultaneous, kashiwamura2022bayesian, ueda2023bayesian, kashiwamura2024noise}. For more details about the EMC algorithm and our simulations, see our GitHub repository described in Data Availability section. 

\subsubsection{Model for estimation}
Let $D \coloneqq \{ t_i, y_i \}_{i=1}^{M}$ represent the observed dataset of the order parameter $y_i$ at time $t_i$, where $M$ is the number of data points. To incorporate the analytical expression of the order parameter dynamics in the thermodynamic limit into the estimation framework, we assume that the observed data $y_i$ can be modeled as the sum of the exact solution~\eqref{eq:oa_sol_0} and the observational noise. Namely, we assume the following relationship: 
\begin{equation} 
\label{eq:data}
y_i = R_{\rm sol}(t_i; \theta) + \xi_i, 
\end{equation} 
where $\xi_i$ denotes the observational noise. For simplicity and analytical tractability, we further assume that $\xi_i$ is subject to a Gaussian
distribution with zero mean and variance $\sigma^2$ [i.e., $\xi_i \sim \mathcal{N}(0, \sigma^2)]$. We set $t_1 = 0$ and $t_M = T$, meaning that the dataset is obtained within the time interval $[0, T]$. For convenience in later analysis, we also introduce the quantity: 
\begin{equation} 
b \coloneqq \frac{1}{\sigma^2}, 
\end{equation} 
which is referred to as the inverse temperature.

\subsubsection{Bayesian formulation}
According to Eq.~\eqref{eq:data}, the conditional probability of the observed data set $D$ for a given set of parameters $\theta$ and noise-variance $1/b$ is calculated as
\begin{align}
    p(D | \theta, b) &= \prod_{i=1}^{M} p(y_i | t_i, \theta, b) \notag \\
    &= \left( \frac{b}{2\pi} \right)^{\frac{M}{2}} \exp \left[ - M b E(\theta) \right], 
    \label{eq:likeli}
\end{align}
where the function $E(\theta)$ given by
\begin{equation}
    E(\theta) \coloneqq \frac{1}{2M} \sum_{i=1}^{M} [y_i - R_{\rm sol}(t_i; \theta)]^2,
    \label{eq:error}
\end{equation}
denotes the error between the observed data $y_i$ and the fitting function $R_{\rm sol}(t_i; \theta)$. 
To derive Eqs.~\eqref{eq:likeli} and \eqref{eq:error}, we use the assumption that the observational noise $\xi_i = y_i - R_{\rm sol}(t_i; \theta)$ in Eq.~\eqref{eq:data} is subject to Gaussian, which implies that 
\begin{align}
    p(y_i | t_i, \theta, b) 
    &= p(y_i - R_{\rm sol}(t_i; \theta)| t_i, \theta, b) \notag \\
    &= \left( \frac{b}{2\pi} \right)^{\frac{1}{2}} \exp \left\{ - \frac{b [y_i - R_{\rm sol}(t_i; \theta)]^2}{2} \right\}.
    \label{eq:likeli_component}
\end{align}

In Bayesian analysis, we treat $\theta$ as a random variable subject to a probability function $p(\theta)$. Here, $p(\theta)$ represents the distribution function of $\theta$ before observing the dataset $D$, which is known as the prior distribution. The choice of prior distribution $p(\theta)$ depends on the specific problem setting. 
Another parameter $b$ is regarded as a hyperparameter and determined by the empirical Bayes method (see Sec.~\ref{sec:para_est} for the details).

By using Bayes' theorem, the posterior distribution of $\theta$ for given $D$ and $b$ is 
\begin{align}
    p(\theta | D, b) &= \frac{p(D, \theta, b)}{\displaystyle \int d \theta p(D, \theta, b)} \notag \\
    &= \frac{p(D| \theta, b) p(\theta)}{\displaystyle \int d \theta p(D| \theta, b) p(\theta)} \notag \\
    &= \frac{1}{Z(b)} \left( \frac{b}{2\pi} \right)^{\frac{M}{2}} \exp \left[ - M b E(\theta) \right] p(\theta). \label{eq:posteri}
\end{align}
Here, the quantity $Z(b)$ is called the marginal likelihood and is given by
\begin{align}
    Z(b) &\coloneqq \int d \theta p(D| \theta, b) p(\theta) \notag \\
    &= \left( \frac{b}{2\pi} \right)^{\frac{M}{2}} \tilde{Z}(b), \label{eq:marginal}
\end{align}
where 
\begin{equation}
    \tilde{Z}(b) \coloneqq \int d\theta \exp \left[ -M b E(\theta) \right] p(\theta).
\end{equation}
We also introduce the Bayesian free energy $F(b)$ as
\begin{equation}
    \label{eq:free}
    F(b) \coloneqq -\log Z(b). 
\end{equation}

\subsubsection{Exchange Monte Carlo method}
\label{sec:method_EMC}
Here we describe the algorithm of the EMC method, by which we numerically obtain the posterior distribution $p(\theta | D, b)$ and the marginal likelihood $Z(b)$.  

We first prepare $L$ different replicas of the system \cite{hukushima1996exchange}. For each of these $L$ replica layers, we associate a corresponding set of model parameters $\{\theta_l \}_{l=1}^L$ and inverse temperatures $\{b_l \}_{l=1}^L$, where the inverse temperatures satisfy $0 = b_1 < b_2 < \cdots < b_L$. In practice, the set of inverse temperatures $\{ b_l \}$ is chosen to follow a geometric series \cite{nagata2008asymptotic, kashiwamura2022bayesian}, i.e., 
\begin{equation}
    \label{eq:geo_series}
    b_l = 
    \begin{cases}
        0 & (l = 1), \\
        Q \eta^{(l - L)} & (l \neq 1), 
    \end{cases}
\end{equation}
where $Q$ and $\eta$ are the hyperparameters of positive real constants. The posterior distribution in each replica is then written as $p(\theta_l | D, b_l)$ for $l = 1, 2, \ldots L$.

In the EMC method, we update $\theta_l$ in each replica such that the joint density 
\begin{equation}
    p(\theta_1, \ldots, \theta_L | D, b_1, \ldots, b_L) \coloneqq \prod_{l=1}^{L} p(\theta_l | D, b_l),
\end{equation}
remains invariant. Here, the initial values of parameters $\{ \theta_1^{(0)}, \ldots, \theta_L^{(0)} \}$ are derived randomly from the prior distribution $p(\theta)$. The algorithms are composed of the following two parts, both of which satisfy the detailed balance condition: 
\begin{enumerate}[label = Step \Roman*.]
    \item Sampling in each replica: we update $\theta_l$ under the probability density $p(\theta_l | D, b_l)$ with a conventional Markov chain Monte Carlo (MCMC) method. The sampling in each replica is performed in parallel. 
    \item Exchange between adjacent replicas: we sequentially exchange the set of model parameters between adjacent replicas (i.e., we exchange $\theta_l$ and $\theta_{l+1}$ for $l = 1, \ldots, L-1$) with the probability $\mu$ given by
    \begin{equation}
        \label{eq:mu}
        \mu \coloneqq \min [1,  \nu],
    \end{equation}
    where
    \begin{align}
        \nu & \coloneqq \frac{p(\theta_{l+1} | D, b_l) p(\theta_l | D, b_{l+1})}{p(\theta_l | D, b_l) p(\theta_{l+1} | D, b_{l+1})} \notag \\
        &= \exp\{M (b_{l+1} - b_l) [E(\theta_{l+1}) - E(\theta_l)]\}.
    \end{align}
    This exchange process is inserted after every $n_{\theta}$ steps of the previous sampling process (step I), where $n_{\theta}$ denotes the number of estimated parameters within the parameter set $\theta$ \cite{iwamitsu2021replica}.
\end{enumerate}
A summary of the above parameter update procedures are provided in Algorithm \ref{alg:EMC}. Here, $S$ denotes the total number of Monte Carlo (MC) steps, that is, the total number of iterations of step I.

By repeating Steps I and II, we obtain the sampling results $\{ \theta_1^{(k)}, \ldots, \theta_L^{(k)} \}_{k=0}^{S}$ from joint probability $p(\theta_1, \ldots, \theta_L | D, b_1, \ldots, b_L)$. The samples from each replica $\{ \theta_1^{(k)}\}_{k=0}^{S}$ is subject to the posterior probability $p(\theta_l | D, b_l)$ \cite{nagata2012bayesian}. In practice, we disregard the sampling results of the first $B$ steps as the burn-in period and only use the rest (i.e., $\{ \theta_1^{(k)}\}_{k=B+1}^{S}$) to estimate the posterior probability.  

\begin{algorithm}[H]
    \caption{Updating parameters in EMC}
    \label{alg:EMC}
    \begin{algorithmic}
        \For{$i=1$ to $S$} 
            \For{$l=1$ to $L$}
                \State Update $\theta_l$ in each replica (Step I).
            \EndFor
            \If{$i/n_\theta$ is an integer}
                \For{$l=1$ to $L-1$}
                    \State Exchange $\theta_l$ and $\theta_{l+1}$ 
                    \State with a probability $\mu$ (Step II).
                \EndFor
            \EndIf
        \EndFor
    \end{algorithmic}
\end{algorithm}

It is important to note that the second algorithm (step II) prevents the trapping in the local minima during the sampling procedure, which is one of the major problems of conventional MCMC methods (step I). 

To perform the Monte Carlo sampling in each replica (step I), we adopt the Metropolis algorithm \cite{metropolis1953equation} with a Gaussian proposal distribution $\mathcal{N}(0, s^2)$. Consulting Ref.~\cite{iwamitsu2021replica}, we update the standard deviation $s$, which can be considered as a step width for the Metropolis sampling, by using the acceptance ratio of Metropolis samplings:
\begin{equation}
    \label{eq:step}
    s \leftarrow 
    \begin{cases}
        [1 + (r - r_{\rm target})] s & {\rm if} \quad |r - r_{\rm target}| > r_{\rm tol}, \\
        s & {\rm otherwise},
    \end{cases} 
\end{equation}
where $r$ denotes the mean acceptance ratio over 200 Metropolis steps. The hyperparameters $r_{\rm target}$ and $r_{\rm tol}$ represent the target acceptance ratio and the tolerance range of the mean acceptance ratio, respectively. 
The update of step width [i.e., Eq.~\eqref{eq:step}] is performed every 200 steps of Metropolis sampling during the burn-in period. 

\subsubsection{Calculation of marginal likelihood}
\label{sec:marginal}
The marginal likelihood $Z(b)$ given by Eq.~\eqref{eq:marginal} can be calculated by the chaining of importance sampling \cite{bishop2006pattern}. Noting that $b_1 = 0$, we have
\begin{align}
    \tilde{Z}(b_l) &= \prod_{k=1}^{l-1} \frac{\tilde{Z}(b_{k+1})}{\tilde{Z}(b_k)} \notag \\
    &= \prod_{k=1}^{l-1} \frac{\displaystyle\int d\theta_{k+1} \exp[-M b_{k+1} E(\theta_{k+1})] p(\theta_{k+1})}{\displaystyle\int d\theta_{k} \exp[-M b_{k} E(\theta_{k})] p(\theta_{k})} \notag \\
    &= \prod_{k=1}^{l-1} \frac{\displaystyle\int d\theta \exp[-M b_{k+1} E(\theta)] p(\theta)}{\displaystyle\int d\theta \exp[-M b_{k} E(\theta)] p(\theta)} \notag \\
    &= \prod_{k=1}^{l-1} \int d\theta \exp[-M (b_{k+1} - b_k) E(\theta)] p(\theta|D, b_k). \label{eq:marginal_chain}
\end{align}
Using the samples $\{ \theta_l \}$ that is subject to $p(\theta_l|D, b_l)$ in each replica, one can numerically calculate the integral in Eq.~\eqref{eq:marginal_chain}. Note that, because the posterior distribution in one replica [i.e., $p(\theta_k|D, b_k)$] is close to that in the adjacent distribution [i.e., $p(\theta_{k+1}|D, b_{k+1})$], it is expected that the numerical calculation of Eq.~\eqref{eq:marginal_chain} is accurate \cite{bishop2006pattern}. \label{rev1_minor_3}
We can also calculate the Bayesian free energy [Eq.~\eqref{eq:free}] as
\begin{equation}
    F(b_l) = -\frac{M}{2} \log \left(\frac{b_l}{2\pi}\right) - \log \tilde{Z}(b_l).
\end{equation}

\subsubsection{Parameter estimation}
\label{sec:para_est}
We adopt the empirical Bayes method (also known as the type II maximum likelihood approach) \cite{bernardo1994bayesian} to estimate the two unknown parameters $\theta$ and $b$. 
We first determine the optimal value of $b_l$, denoted by $b_{\hat{l}}$, by maximizing $p(D|b_l)$, the likelihood function with respect to $b_l$,
i.e., 
\begin{align}
    \hat{l} 
    &= \argmax_l\: p(D|b_l) \notag \\
    &= \argmax_l \int d \theta p(D|\theta, b_l) p(\theta) \notag \\
    &= \argmax_l Z(b_l) \label{eq:inv_opt} \\
    &= \argmin_l F(b_l). \label{eq:inv_opt_f}
\end{align}
In other words, we select $b_{\hat{l}}$ such that it maximizes the marginal likelihood $Z(b_l)$. 
Of note, the optimal value of $b_{l}$ can also be obtained by the hierarchical Bayes approach \cite{gelman1995bayesian, tokuda2017simultaneous} (see Appendix~\ref{sec:app_hier}).

Using the optimal $b_{\hat{l}}$ obtained from Eq.~\eqref{eq:inv_opt_f}, we estimate the model parameter $\theta$ via the maximum a posteriori (MAP) estimation: 
\begin{equation}
    \label{eq:MAP}
    \hat{\theta} = \argmax_{\theta} p(\theta | D, b_{\hat{l}}),
\end{equation}
where $\hat{\theta}$ denotes the the MAP estimator of $\theta$. 

\section{Numerical experiments}
\label{sec:result}
To confirm the validity of our framework, we first use the same model for both data generation and estimation. Namely, we perform the EMC method using the artificial dataset generated from the analytical solution~\eqref{eq:oa_sol_0}. The results of the first experiment are shown in Sec.~\ref{sec:res_oa}. Then, we use the Kuramoto model~\eqref{eq:kura} as the data generation model to investigate the finite size effect of the number of oscillators. The results of the second experiment are shown in Sec.~\ref{sec:res_kura}. 

Before we show the results of the EMC estimation, we summarize the setups for the numerical experiments.

\subsection{Experimental setups}
\label{sec:setting}
\subsubsection{Property of time series data}
In the present study, we focus on the desynchronizing process from the complete synchrony state (i.e., $R_0 = 1$), motivated by several experiments in plant biology where oscillating units (e.g., the circadian clock in cells) desynchronize without the entrainment by the light-dark cycle \cite{wenden2012spontaneous, muranaka2016heterogeneity, watanabe2023non}. Thus, we use the time series data where the order parameter monotonically decreases. 

\subsubsection{Assumption on model parameter and initial condition}
Since we address the desynchronizing process, we assume 
\begin{equation}
    \label{eq:lam_posi}
    \lambda > 0, 
\end{equation}
so that the order parameter $R(t)$ that follows Eq.~\eqref{eq:oa} decreases with time. 

We also assume that the initial value of the order parameter ($R_0 = 1$) is given. 
Under this assumption, we can rewrite the set of model parameters $\theta$, which is originally given in Eq.~\eqref{eq:theta}, as 
\begin{equation}
    \label{eq:theta_new}
    \theta = \{ \gamma, \lambda \}.
\end{equation} 
The analytical solution~\eqref{eq:oa_sol_0} can also be rewritten as 
\begin{equation}
    \label{eq:oa_sol}
    R_{\rm sol}(t; \theta) = \frac{e^{-\lambda t}}{\sqrt{1 + \frac{\lambda - \gamma}{\lambda}\left( e^{-2 \lambda t} - 1 \right)}}.
\end{equation}
In the numerical experiments in Secs.~\ref{sec:res_oa} and \ref{sec:res_kura}, we use $R_{\rm sol}(t; \theta)$ in Eq.~\eqref{eq:oa_sol} as the exact solution and estimate the set of parameters $\theta$ in Eq.~\eqref{eq:theta_new}. We reconstruct the original model parameter $K$ as $K = 2(\gamma - \lambda)$ during the estimation process.

\subsubsection{Assumption on prior distribution}
We assume that the prior probability $p(\theta)$ can be written as the product of the prior probabilities of each parameter, i.e., 
\begin{equation}
        p(\theta) = p(\gamma) p(\lambda). 
\end{equation}
We also assume that both of $p(\gamma)$ and $p(\lambda)$ are subject to uniform distribution, i.e., 
\begin{align}
    p(\gamma) &=
    \begin{cases}
        \frac{1}{v_{\gamma} - u_{\gamma}} & {\rm if} \quad u_{\gamma} < \gamma < v_{\gamma}, \\
        0 & {\rm otherwise},
    \end{cases} \\
    p(\lambda) &=
    \begin{cases}
        \frac{1}{v_{\lambda} - u_{\lambda}} & {\rm if} \quad u_{\lambda} < \lambda < v_{\lambda}, \\
        0 & {\rm otherwise}, 
    \end{cases} 
\end{align}
where $u_{\gamma}, v_{\gamma}, u_{\lambda}$, and $v_{\lambda}$ are hyperparameters. According to Eq.~\eqref{eq:lorentz} and the assumption~\eqref{eq:lam_posi}, we set $u_{\gamma} \geq 0$ and $u_{\lambda} \geq 0$. 

\subsubsection{Values of fixed parameters}
When generating the artificial datasets, we use the different values of noise strength $\sigma$ and oscillator number $N$ in the first (Sec.~\ref{sec:res_oa}) and second (Sec.~\ref{sec:res_kura}) experiments, respectively. We fix the values of the remaining parameters, which are summarized in Table~\ref{tab:para}. 

\begin{table}
\caption{Parameters used for generating datasets and performing EMC \label{tab:para}}
\begin{ruledtabular}
\begin{tabular}{cll}
Parameter & Value & Meaning \\ 
\hline
$K$ & 0.05 & coupling strength \\
$\gamma$ & 0.08 & heterogeneity of oscillators \\
$M$ & 101 & number of data points \\
$T$ & 50 & maximum observation time \\
$L$ & 50 & number of replicas \\
$Q$ & 1500000 & maximum inverse temperature \\
$\eta$ & 1.5 & ratio of adjacent inverse temperatures \\
$u_{\gamma}$ & 0.0 & lower bound of $p(\gamma)$ \\
$v_{\gamma}$ & 1.0 & upper bound of $p(\gamma)$ \\
$u_{\lambda}$ & 0.0 & lower bound of $p(\lambda)$ \\
$v_{\lambda}$ & 1.0 & upper bound of $p(\lambda)$ \\
$S$ & 100000 & number of total Metropolis steps \\
$B$ & 50000 & number of steps within burn-in period \\
$r_{\rm target}$ & 0.6 & target acceptance ratio \\
$r_{\rm tol}$ & 0.05 & tolerance range of acceptance ratio \\
$n_{\theta}$ & 2 & number of estimated parameters
\end{tabular}
\end{ruledtabular}
\end{table}

\subsection{Estimation from data generated by Ott-Antonsen formula}
\label{sec:res_oa}
To confirm the validity of our framework, we inversely estimate the model parameter $\theta$ from the artificial data generated by adding an observation noise to the analytical solution~\eqref{eq:oa_sol}. Namely, we first create the dataset $D$ according to Eq.~\eqref{eq:data} with different noise strength $\sigma$, and then we estimate the model parameters $\theta$ using the framework described in Sec.~\ref{sec:frame}. 

\subsubsection{Typical estimation results}
\label{sec:oa_typ}
\begin{figure*}
    \centering
    \includegraphics[width = .94\linewidth]{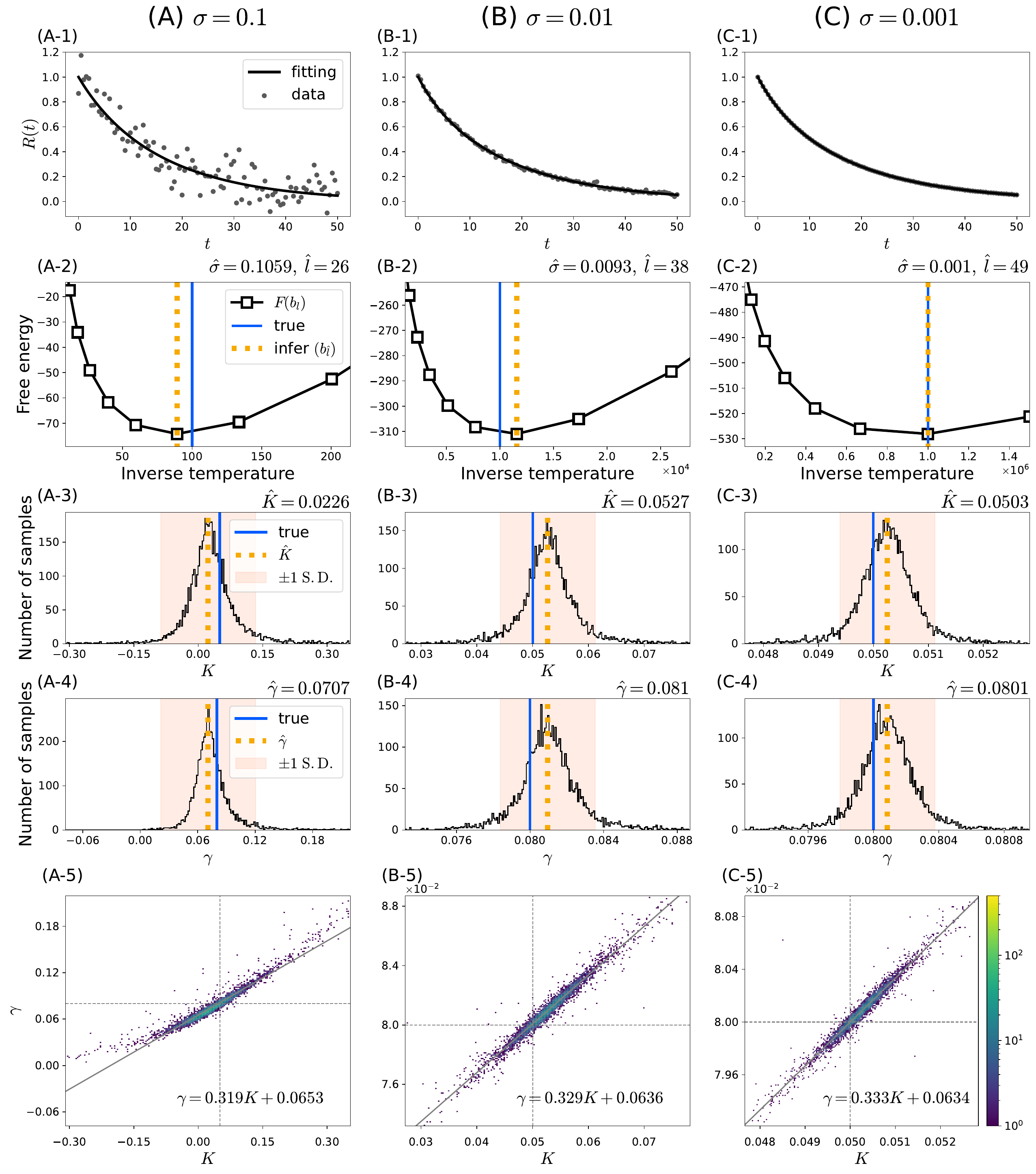}
    \caption{The EMC estimation results from the datasets generated from Eq.~\eqref{eq:data}. For each column, we create a single dataset with different noise intensities [i.e., $\sigma = 0.1,\, 0.01$, and $0.001$ in columns (A), (B), and (C), respectively] and perform the EMC simulation once. First row: the generated dataset (scatter plots) and the exact solution calculated with the MAP estimators $R_{\rm sol}(t; \hat{\theta})$ (black solid line). Second row: the Bayesian free energy $F(b_l)$ at inverse temperatures $b_l$ (the black square marks that are connected linearly). The solid and dotted vertical lines show the inverse of the true noise variance $1/\sigma^2$ and the inferred inverse temperature $b_{\hat{l}}$, respectively. At the top right of each panel, we show the values of the inferred index $\hat{l}$ and inferred noise strength $\hat{\sigma} \coloneqq 1/b_{\hat{l}}^2$. 
    Third and fourth rows: histograms of the estimated posterior distribution of $K$ and $\gamma$, respectively. The light red shaded area in each panel represents the interval of $\pm$ 1 standard deviation (SD) of the posterior distribution around the MAP estimator. The solid and dotted vertical lines denote the true value and the MAP estimator, respectively. Fifth row: two-dimensional histograms of the estimated posterior distributions, with corresponding linear regression results overlaid as gray lines. The color bar indicates the number of samples contained within each bin.
    The equation of each regression line is displayed in the lower right corner of the respective panel. 
    Black dotted lines indicate the true parameter values. For all histograms, the parameter ranges for $K$ and $\gamma$ are set to the MAP estimator $\pm$ three SDs of the posterior distribution, and each range is divided into 200 bins. Linear regression is performed using EMC samples within the histogram range to exclude outliers. 
    }
    \label{fig:OA_single}
\end{figure*}

Figure~\ref{fig:OA_single} shows the results of the EMC simulation performed under different noise strengths $\sigma$. Each column in the figure corresponds to the estimation results using a dataset generated from Eq.~\eqref{eq:data}, with noise strength set to $\sigma = 0.1$ [column (A)], $\sigma = 0.01$ [column (B)], and $\sigma = 0.001$ [column (C)]. The EMC simulation is performed once for each dataset displayed in the top row of Fig.~\ref{fig:OA_single} [panels (A-1), (B-1) and (C-1)] as scatter plots. Of note, the convergence of each EMC simulation is evaluated in Fig.~\ref{fig:OA_error} in Appendix~\ref{sec:app_err}.

In the second row of Fig.~\ref{fig:OA_single}, we plot the Bayesian free energy $F(b_l)$ against different values of inverse temperatures $b_l$. The optimal replica index $\hat{l}$ and corresponding inverse temperature $b_{\hat{l}}$ are determined by minimizing the free energy, as described in Eq.~\eqref{eq:inv_opt_f}. 
\label{page:inv_temp_disc}
Although the inferred inverse temperature (dotted vertical line) closely approximates the true value $1/\sigma^2$ (solid vertical line), they do not coincide exactly in panels (A-2) and (B-2) of Fig.~\ref{fig:OA_single}. This discrepancy arises because the inverse temperature $b$ is selected from a discrete set $\{b_l\}_{l=1}^{L}$, which does not include the true values corresponding to $\sigma = 0.1$ or $0.01$. We expect that increasing the number of candidate values (i.e., increasing the total number of replicas $L$) would lead to more accurate estimation of noise intensity, even though this would also increase the computational cost. 

The third and fourth rows of Fig.~\ref{fig:OA_single} display the estimated posterior distributions of the parameters $K$ and $\gamma$, respectively. These distributions are obtained as histograms of the Monte-Carlo samples from the $\hat{l}$-th replica. Examining the estimated values $\hat{K}$ and $\hat{\gamma}$ described in the top-right corner of each panel, we observe that the accuracy of the estimates deteriorates as the strength of observational noise increases from right to left (note that the ranges of the horizontal axes differ across panels). Nonetheless, in all cases, the true parameter values remain within the $\pm 1$ standard deviation (SD) interval of the posterior distribution, which is indicated by the red shaded region.

\label{page:corr_OA}
The fifth row of Fig.~\ref{fig:OA_single} represents the two-dimensional histograms, which are estimates of posterior distributions in two-dimensional parameter space. 
These histograms reveal a strong positive correlation between the estimated $K$ and $\gamma$ samples. Notably, the EMC samples from different datasets -- each generated under varying observational noise intensities -- lie approximately along the same regression line, given by
\begin{equation}
\label{eq:corr}
    \gamma = 0.33 K + 0.0635.
\end{equation}
We observe that varying $K$ and $\gamma$ along this regression line~\eqref{eq:corr} does not significantly alter the shape of the analytical solution~\eqref{eq:oa_sol} (see Fig.~\ref{fig:correlations}), and consequently, has little effect on the estimation error~\eqref{eq:error}.
A more detailed discussion and analysis of this correlation are provided in Appendix~\ref{sec:app_corr}.

\subsubsection{Dependence of estimation accuracy on noise intensity}
\begin{figure*}
    \centering
    \includegraphics[width = .9\linewidth]{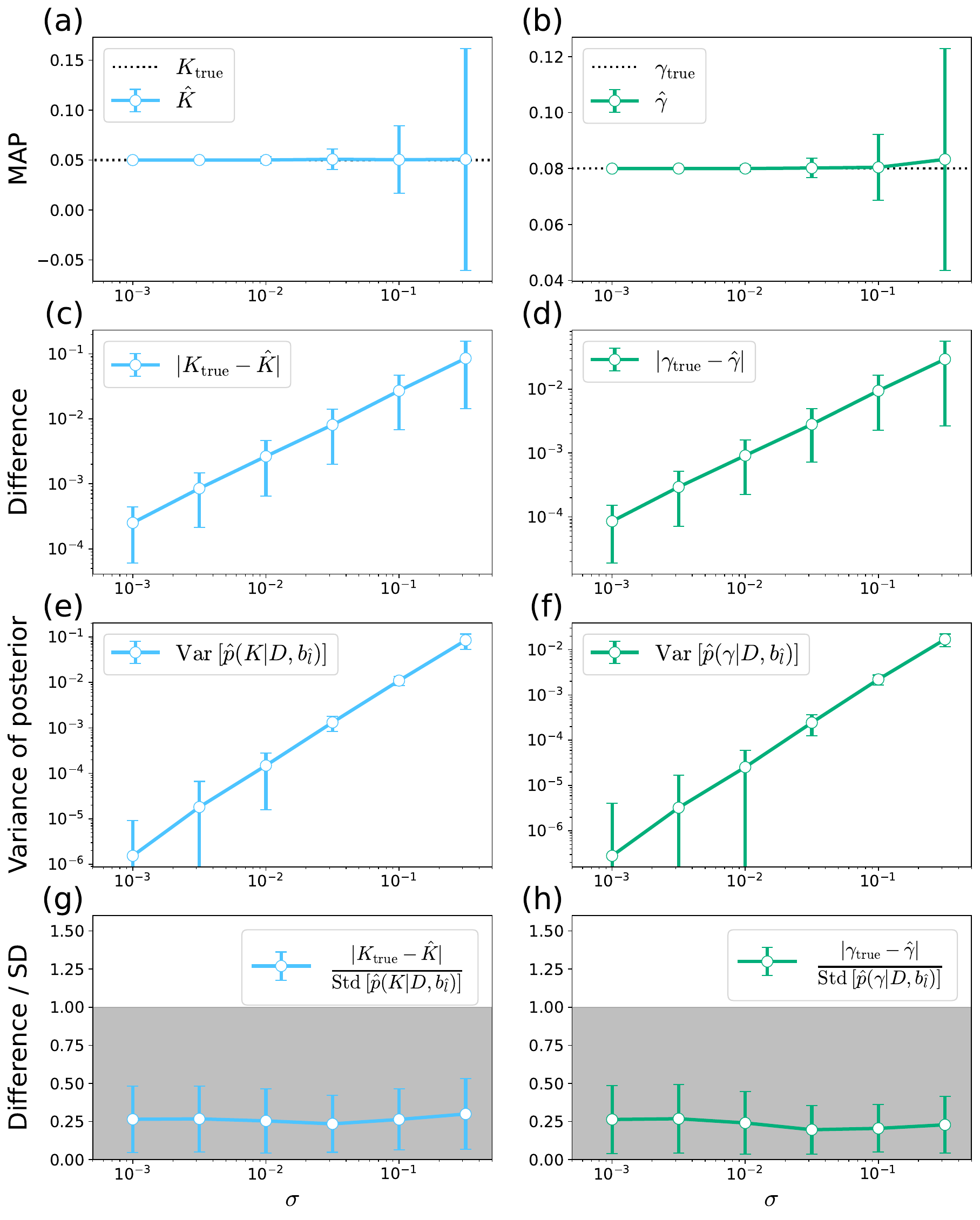}
    \caption{Dependence of inference accuracy on the noise intensity $\sigma$. For each $\sigma$, the EMC simulations are repeated using 1000 datasets generated from Eq.~\eqref{eq:data} with different random seeds. Panels (a) and (b): the means and standard deviations of the MAP estimators $\hat{K}$ and $\hat{\gamma}$, respectively. The black dotted lines represent the true parameter values. Panels (c) and (d): the mean and the standard deviation of the absolute error between the MAP estimates and the true parameter values. Panels (e) and (f): the means and standard deviations of the variances of the posterior distributions for $K$ and $\gamma$. The notations $\hat{p} (K|D, b_l)$ and $\hat{p} (\gamma|D, b_l)$ in the legends denote the estimated posterior distribution for $K$ and $\gamma$, respectively. Panels (g) and (h): the ratios of the absolute estimation error to the standard deviation of the corresponding posterior distribution. Grey bands indicate the range $[0,1]$, where the true parameter lies within one SD of the posterior distribution centered at the MAP estimator.}
    \label{fig:OA_rep}
\end{figure*}

To quantitatively examine the effect of noise strength on estimation results, we repeat EMC simulations on datasets generated with six different values of noise intensity ($\sigma$), as shown in Fig.~\ref{fig:OA_rep}. For each $\sigma$, we generate 1000 independent datasets from Eq.~\eqref{eq:data} using different random seeds, and perform the EMC simulation on each dataset. Various statistical measures are computed and compared across the different noise levels to evaluate how inference performance depends on the observational noise.

\label{rev:loglog1}
Figures~\ref{fig:OA_rep} (a) and (b) display the mean and standard deviation of the MAP estimators obtained from repeated EMC simulations. These results indicate that, though the estimates exhibit increasing variability with larger noise intensities, the mean remains close to the true parameter values. Figures~\ref{fig:OA_rep} (c) and (d) present the absolute estimation error, revealing the convergence of the MAP estimators toward the true values as the noise intensity decreases. 

To assess how the posterior distributions change with noise intensity, we compute the variances of the estimated posterior distributions from each EMC simulation and plot their means and standard deviations in Figs.~\ref{fig:OA_rep} (e) and (f). These plots demonstrate that the posterior distributions become more concentrated as $\sigma$ decreases.

Finally, Figs.~\ref{fig:OA_rep} (g) and (h) show the ratios of the absolute estimation error to the standard deviation of the corresponding posterior distribution. These ratios remain within the interval $[0,1]$ on average, regardless of the noise strength, indicating that the true parameter values typically lie within one standard deviation of the posterior distribution. This observation reflects the fact that the same model [Eq.~\eqref{eq:data}] is used for both data generation and parameter inference.

\subsection{Estimation from data generated by Kuramoto model}
\label{sec:res_kura}
Next, we infer the model parameters $\theta$ using the time series data generated from the Kuramoto model~\eqref{eq:kura} with the Lorentzian distribution of natural frequencies [Eq.~\eqref{eq:lorentz}]. When creating the time series of the order parameter, we numerically integrate Eq.~\eqref{eq:kura} by the fourth-order Runge-Kutta method with the time step $0.001$ and calculate the Kuramoto order parameter $R(t)$ for each data point. For the initial condition, we fix $\phi_i(0)=0$ for all $i$, which corresponds to $R(0)=1$.

\subsubsection{Typical estimation results}
\label{sec:kura_typ}
\begin{figure*}
    \centering
    \includegraphics[width = .94\linewidth]{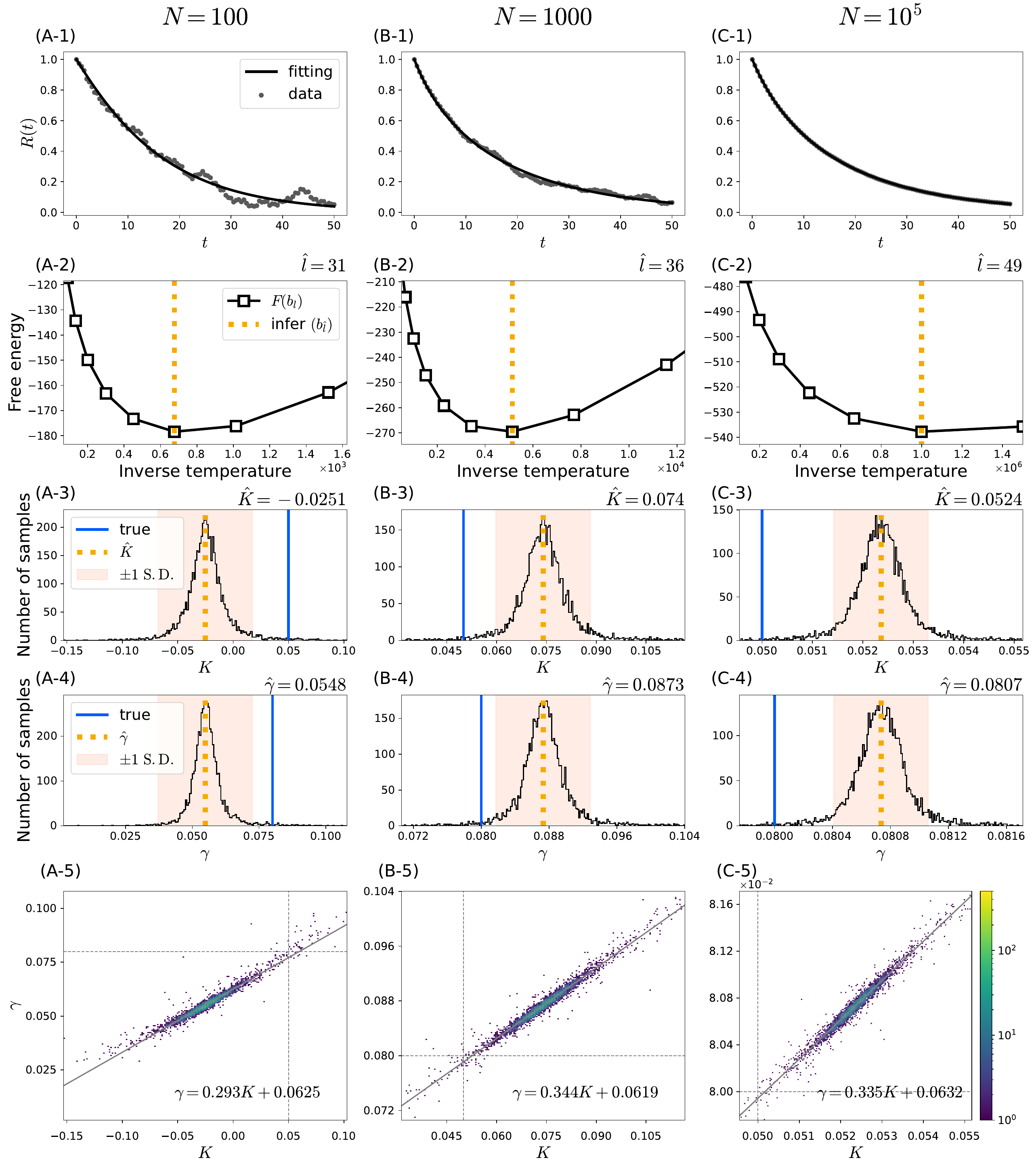}
    \caption{The EMC estimation results from the datasets generated from the Kuramoto model~\eqref{eq:kura}. For each column, we create a single dataset with different numbers of oscillators [i.e., $N = 100,\, 1000$, and $10^5$ in columns (A), (B), and (C), respectively] and perform the EMC simulation once. First row: the generated dataset (scatter plots) and the exact solution with the MAP values $R_{\rm sol}(t; \hat{K}, \hat{\gamma})$. Second row: the Bayesian free energy $F(b_l)$ at inverse temperatures $b_l$ (the black square marks that are connected linearly). The dotted vertical lines show the inferred inverse temperature $b_{\hat{l}}$. Third and fourth rows: histograms of the estimated posterior distribution of $K$ and $\gamma$, respectively. 
    The light red shaded area in each panel represents the interval of $\pm$ 1 SD of the posterior distribution around the MAP estimator. The solid and dotted vertical lines denote the true value and the inferred MAP estimator, respectively. Fifth row: two-dimensional histograms of the estimated posterior distributions, with corresponding linear regression results overlaid as gray lines. The color bar indicates the number of samples contained within each bin.
    The equation of each regression line is displayed in the lower right corner of the respective panel. 
    Black dotted lines indicate the true parameter values. For all histograms, the parameter ranges for $K$ and $\gamma$ are set to the MAP estimator $\pm$ three SDs of the posterior distribution, and each range is divided into 200 bins. Linear regression is performed using EMC samples within the histogram range to exclude outliers.}
    \label{fig:kura_single}
\end{figure*}

Figure~\ref{fig:kura_single} presents the estimation results for three different datasets. The datasets used in columns (A), (B), and (C) are generated from Eq.~\eqref{eq:kura} with $N=100$, $1000$, and $10^5$, respectively, and are shown as scatter plots in the first row of Fig.~\ref{fig:kura_single} [panels (A-1), (B-1) and (C-1)]. For each dataset, a single EMC simulation is performed. The convergence of each EMC simulation is evaluated in Fig.~\ref{fig:kura_error} in Appendix~\ref{sec:app_err}.

In the second row of Fig.\ \ref{fig:kura_single}, we plot the Bayesian free energy $F(b_l)$ as a function of the inverse temperature $b_l$, and identify the optimal replica index $\hat{l}$ that minimizes the free energy. 
The third and fourth rows of Fig.~\ref{fig:kura_single} present the estimated posterior distributions for $K$ and $\gamma$, respectively, shown as histograms of the Monte-Carlo samples from $\hat{l}$-th replica. As $N$ increases from left to right, the MAP estimators $\hat{K}$ and $\hat{\gamma}$, indicated in the top-right corner of each panel, approach the true values. However, in contrast to Fig.~\ref{fig:OA_single}, the true parameter values fall outside the $\pm 1$ SD of the posterior distribution even when the MAP estimators are close to the true values, as observed in the case when $N=10^5$.

The fifth row of Fig.~\ref{fig:kura_single} represents the estimated posterior distributions in the two-dimensional parameter space. 
\label{page:corr_kura}
As in the previous section (Sec.~\ref{sec:res_oa}), the two-dimensional histograms reveal a strong positive correlation between the estimated $K$ and $\gamma$ samples. These samples lie approximately along the same regression line described in Eq.~\eqref{eq:corr}. A detailed discussion and analysis of this correlation is provided in Appendix~\ref{sec:app_corr}.

\subsubsection{Dependence of estimation accuracy on the number of oscillators}
\begin{figure*}
    \centering
    \includegraphics[width = .9\linewidth]{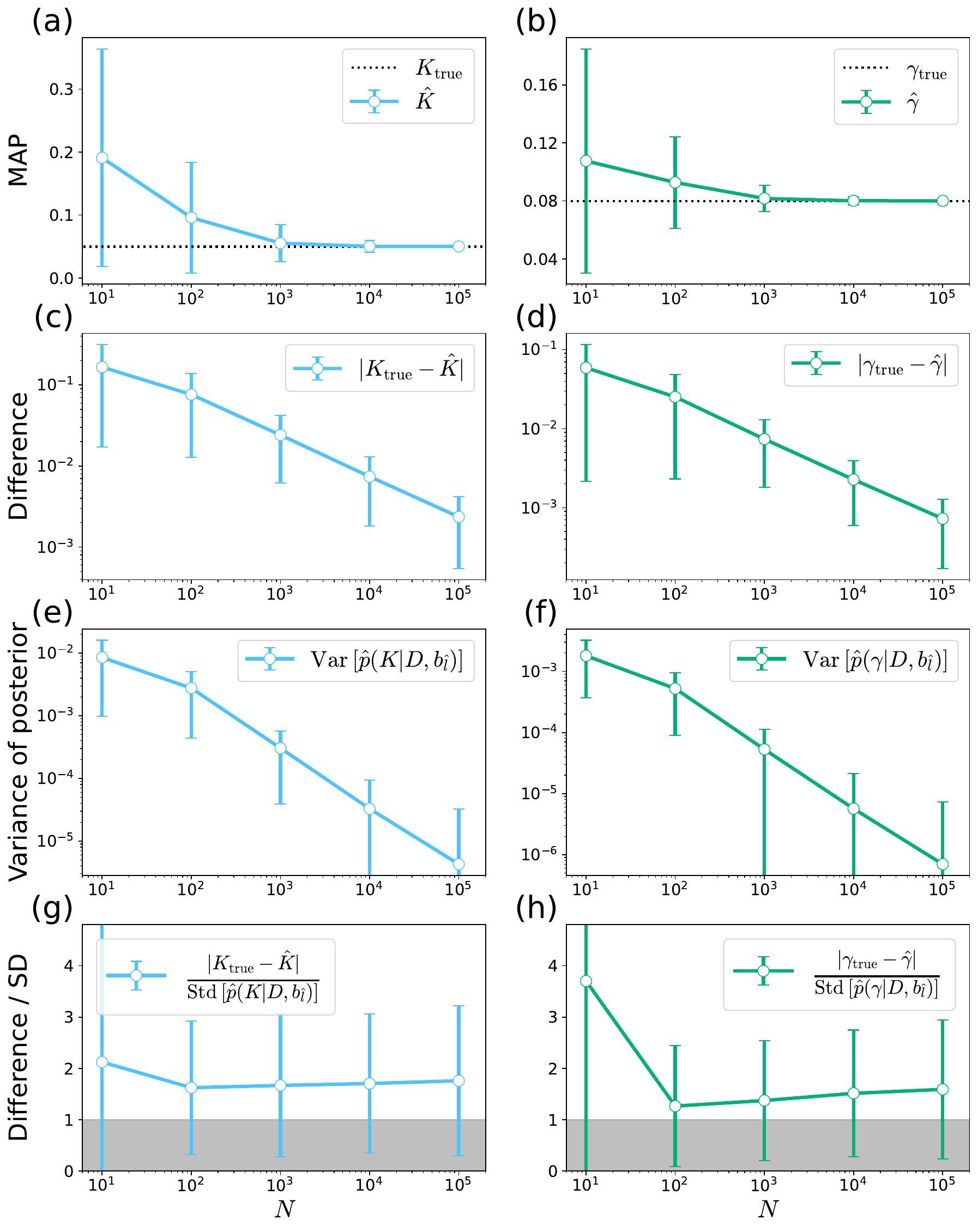}
    \caption{Dependence of inference accuracy on the number of oscillators $N$. For each value of $N$, the EMC simulations are repeated using 1000 datasets generated from Eq.~\eqref{eq:kura} with different random seeds. Panels (a) and (b): the means and standard deviations of the MAP estimators $\hat{K}$ and $\hat{\gamma}$, respectively. The black dotted lines represent the true parameter values. Panels (c) and (d): the mean and the standard deviation of the absolute error between the MAP estimates and the true parameter values. Panels (e) and (f): the means and standard deviations of the variances of the posterior distributions for $K$ and $\gamma$. Panels (g) and (h): the ratios of the absolute estimation error to the standard deviation of the corresponding posterior distribution. Grey bands indicate the range $[0,1]$, where the true parameter lies within one SD of the posterior distribution centered at the MAP estimator.}
    \label{fig:kura_rep}
\end{figure*}

To quantitatively examine how the oscillator number $N$ affects the estimation results, we repeat EMC simulations on datasets generated with five different values of $N$, as shown in Fig.~\ref{fig:kura_rep}. For each $N$, we generate 1000 independent datasets from Eq.~\ref{eq:kura} using different random seeds, and perform the EMC simulation on each dataset. We compute various statistical quantities for each oscillator number and compare the results across different values of $N$. 

Figures~\ref{fig:kura_rep} (a) and (b) show the mean and standard deviation of the MAP estimators for each value of $N$. These results suggest that the MAP estimators become more accurate and exhibit reduced fluctuations as $N$ increases. In contrast to Figs.~\ref{fig:OA_rep} (a) and (b), where the MAP estimators fluctuate around the true values for large $\sigma$, the estimators for small $N$ in Figs.~\ref{fig:kura_rep} (a) and (b) are consistently biased toward larger values. Figure~\ref{fig:kura_rep} (c) and (d) show the mean and standard deviation of the absolute estimation error, suggesting the convergence of the MAP estimators toward the true values as the system size $N$ increases. \label{rev:loglog2} In Figs.~\ref{fig:kura_rep} (e) and (f), we compare the variances of the estimated posterior distributions for different values of $N$. These panels show that the posterior distributions become sharper and exhibit less variability across EMC simulations as $N$ increases.

Figure~\ref{fig:kura_rep} (g) and (h) show the ratios of the absolute estimation error to the standard deviation of the corresponding posterior distribution. These ratios exceed $1$ on average across all values of $N$, indicating that the true parameter values typically fall outside one standard deviation of the posterior distribution. This behavior is in contrast to the results shown in Fig.~\ref{fig:OA_rep} (g) and (h), and reflects a mismatch between the data generation model [Eq.~\eqref{eq:kura}] and the model used for estimation [Eq.~\eqref{eq:data}].

Of note, the datasets shown in Figs.~\ref{fig:OA_single} and \ref{fig:kura_single} are selected based on the results of the multiple EMC simulations presented in Figs.~\ref{fig:OA_rep} and \ref{fig:kura_rep}, respectively. We choose the datasets in Figs.~\ref{fig:OA_single} and \ref{fig:kura_single} so that the difference between the inferred MAP estimator and the true value is closest to the corresponding mean absolute error among the $1000$ EMC simulations, as indicated by the white dot in panels (c) and (d) of Figs.~\ref{fig:OA_rep} and \ref{fig:kura_rep}. For more details about the simulation codes, see our GitHub repository described in Data Availability section. 

\subsubsection{Difference between Kuramoto model dynamics and analytical solution}
\label{sec:ACF}
\begin{figure}
    \includegraphics[width=.98\linewidth]{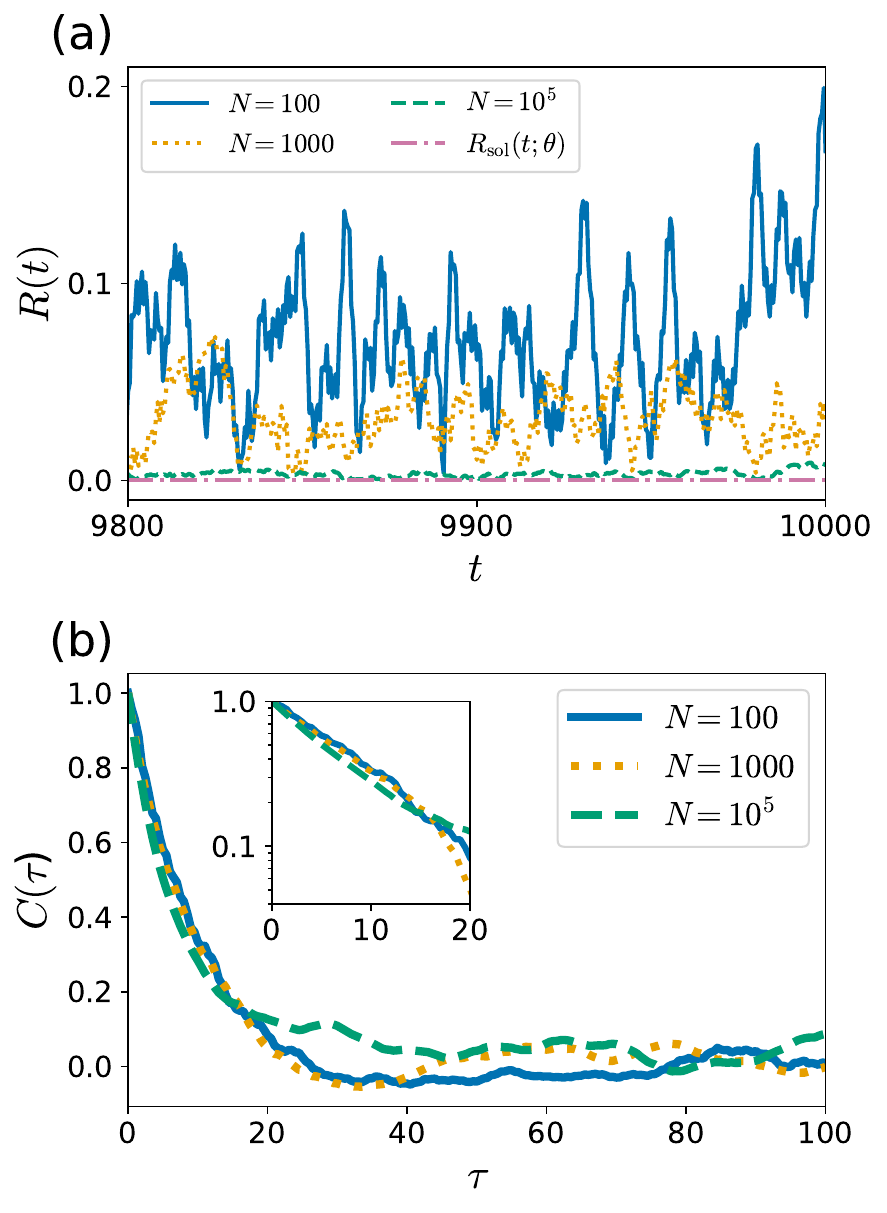}%
    \caption{(a) Long-time behavior of the order parameter. For different system sizes $N$, we numerically integrate the Kuramoto model~\eqref{eq:kura} up to $t=10000$, compute the Kuramoto order parameter $R(t)$ at intervals of $\Delta t = 0.5$, and plot the last segment of the trajectory over $t = 9800$ to $10000$. The solid, dotted, and dashed lines correspond to the case when $N=100, 1000$, and $10^5$, respectively. The analytical solution $R_{\mathrm{sol}}(t;\theta)$ is shown as a dash-dotted line. Simulation parameters, random seeds, and initial conditions are the same as those used for data generation in Fig.~\ref{fig:kura_single}. (b) Autocorrelation functions of the difference between the numerically obtained order parameter $R(t)$ and the analytical solution $R_{\mathrm{sol}}(t;\theta)$. Using the latter half of the time series [i.e., $t \in [5000, 10000]$], we calculate the autocorrelation function of the difference $R(t) - R_{\mathrm{sol}}(t;\theta)$, as defined in Eq.~\eqref{eq:ACF}. The inset shows a log-scale plot highlighting the behavior at small time lags $\tau$.}
    \label{fig:long_data}
\end{figure}

As shown in Figs.~\ref{fig:kura_single} (C-3), (D-3) and Figs.~\ref{fig:kura_rep} (g), (h), the true parameter values lie outside the $\pm 1$ SD range of the corresponding posterior disributions, even for the large system sizes (e.g., $N = 10^5$). This observation suggests that the dynamics of the order parameter in the Kuramoto model~\eqref{eq:kura} cannot be fully captured by a model consisting of the analytical solution plus white Gaussian noise [i.e., Eq.~\eqref{eq:data}]. To further investigate the complex behavior of the order parameter, we generate sufficiently long time series by numerically integrating the Kuramoto model~\eqref{eq:kura} and compute the autocorrelation functions (Fig.~\ref{fig:long_data}). 

Figure~\ref{fig:long_data} (a) shows the long-time behavior of the order parameter of the Kuramoto model~\eqref{eq:kura}, simulated up to $t=10000$. Assuming that the system reaches stationarity after $t = 5000$, we calculate the autocorrelation function of the deviation between the numerically obtained order parameter $R(t)$ and the analytical solution~$R_{\mathrm{sol}}(t;\theta)$, defined as
\begin{equation}
    \label{eq:ACF}
    C(\tau) \coloneqq \frac{\displaystyle\frac{1}{n-\frac{\tau}{\Delta t}}\sum_{i=1}^{n-\frac{\tau}{\Delta t}}[R_{\mathrm{dif}}(t_i)-\mu_{R}] [R_{\mathrm{dif}}(t_i + \tau)-\mu_{R}]}{\sigma_{R}^2},
\end{equation}
where
\begin{align}
    R_{\mathrm{dif}}(t) &\coloneqq R(t) - R_{\mathrm{sol}}(t;\theta), \label{eq:R_dif} \\
    \mu_{R} &\coloneqq \frac{1}{n} \sum_{i=1}^{n}R_{\mathrm{dif}}(t_i), \\
    \sigma_{R}^2 &\coloneqq \frac{1}{n} \sum_{i=1}^{n}[R_{\mathrm{dif}}(t_i)-\mu_{R}]^2.
\end{align}
Here, $\{t_i\}$ denotes the set of time points in the interval $t_1 = 5000$ to $t_n = 10000$, and $\Delta t \coloneqq t_{i+1} - t_i$ is the sampling interval.

Figure~\ref{fig:long_data} (b) shows the autocorrelation functions [Eq.~\eqref{eq:ACF}] computed for different values of $N$. We observe a gradual decay of positive autocorrelation up to $\tau = 20$; for larger lags, the autocorrelation fluctuates around zero. This behavior is inconsistent with the white Gaussian noise assumption in Eq.~\eqref{eq:data}, which implies a delta-function autocorrelation. The inset log-plot in Fig.~\ref{fig:long_data} (b) shows that the autocorrelation decays approximately exponentially for small time lags $\tau$, suggesting that the deviation $R_{\mathrm{dif}}(t)$ may be well approximated by an Ornstein–Uhlenbeck (OU) process, which is characterized by an exactly exponential autocorrelation function \cite{gardiner2009stochastic}.

\section{Discussion and conclusion}
\label{sec:dis}
In this study, we propose a Bayesian framework for estimating the parameters of the Kuramoto model from time series data of the order parameter. After outlining the problem settings and describing the parameter estimation procedure based on EMC simulation in Sec.~\ref{sec:pre}, we evaluate the validity and accuracy of the proposed method through two different numerical experiments. 

\label{rev1_6_begin}
In Sec.~\ref{sec:res_oa}, we examine a benchmark case in which the data generation and estimation models are identical [Eq.~\eqref{eq:data}]. Under this setting, we observe that the MAP estimators converge toward the true parameter values as the noise intensity $\sigma$ decreases [Figs.~\ref{fig:OA_single} (A-3)--(C-3), (A-4)--(C-4) and Figs.~\ref{fig:OA_rep} (c), (d)]. In addition, the posterior distributions become narrower with decreasing noise intensities [Figs.~\ref{fig:OA_rep} (e) and (f)], while including the true parameter values within their $\pm 1$ SD around the corresponding MAP estimators [Figs.~\ref{fig:OA_rep} (g) and (h)]. These results demonstrate the internal consistency of our framework and suggest that the EMC simulations are properly implemented.

We next apply our framework to datasets generated from the Kuramoto model [Eq.~\eqref{eq:kura}], as described in Sec.~\ref{sec:res_kura}. As shown in Figs.~\ref{fig:kura_rep} (c) and (d), the MAP estimators converge toward the true parameter values as the system size $N$ increases, suggesting that our framework remains effective for the parameter inference of the large-scale Kuramoto model. However, in contrast to the benchmark case in Sec.~\ref{sec:res_oa}, the absolute estimation error consistently exceeds one SD of the corresponding posterior distribution, even for large $N$ [Figs.~\ref{fig:kura_rep} (g) and (h)]. This observation reflects a fundamental discrepancy between the data-generating model [Eq.~\eqref{eq:kura}] and the estimation model [Eq.~\eqref{eq:data}]. Thus, we interpret the improved inference accuracy at large $N$ [shown in Figs.~\ref{fig:kura_rep} (c) and (d)] not as an evidence that the estimation model [Eq.~\eqref{eq:data}] fully captures the dynamics of the Kuramoto order parameter, but rather as a consequence of the analytical solution [Eq.~\eqref{eq:oa_sol}] becoming increasingly accurate in the thermodynamic limit ($N \to \infty$).

In search of a more suitable estimation model, we numerically investigate the long-time behavior of the order parameter and compute the autocorrelation function of its deviation from the analytical solution. Figure~\ref{fig:long_data} (b) shows that the autocorrelation function exhibits an initial exponential decay for different system sizes $N$, indicating that the deviation $R_{\mathrm{dif}}(t)$ possesses temporal correlations resembling those of an OU process. These findings are partially consistent with a recent study that employs a two-dimensional OU process for stochastic model reduction of the finite-size Kuramoto model \cite{yue2024stochastic}. Notably, our results in Fig.~\ref{fig:long_data} (b) suggest that the OU-like temporal correlation persist even at large system sizes (e.g., $N=10^5$). Based on this observation, we expect that estimation accuracy could be further improved by replacing the white Gaussian noise in the current estimation model [Eq.~\eqref{eq:data}] with an OU process, yielding the following model:
\begin{equation}
    \label{eq:data2}
    y_i = R_{\rm sol}(t_i; \theta) + x(t_i),
\end{equation}
where $x(t)$ evolves according to the OU process
\begin{equation}
    \dot x = - \nu x + \xi(t).
\end{equation}
with $\nu > 0$ and $\xi(t)$ denoting white Gaussian noise satisfying $\langle \xi(t) \xi(t') \rangle = \sigma^2 \delta(t - t')$. Incorporating the modified estimation model~\eqref{eq:data2} into our Bayesian framework and evaluating its performance remain important directions for future research.
\label{rev1_6_end}

As we describe in Sec.~\ref{sec:marginal}, the marginal likelihood can be accurately calculated by using the EMC method. Since the marginal likelihood represents the validity of the model, we can expand our estimation framework to compare multiple coupled oscillator models. For example, we can also consider the Sakaguchi-Kuramoto model \cite{sakaguchi1986soluble}, which is a generalization of the Kuramoto model \eqref{eq:kura}
with an additional parameter $\alpha$ that denotes the phase shift. By comparing the marginal likelihood of the Kuramoto model~\eqref{eq:kura} and the Sakaguchi-Kuramoto model, we can estimate which model is more likely for a given dataset. The model comparison of different coupled oscillator systems is another important future direction, which can be especially useful when selecting an appropriate model based on experimental data. 

\label{pg:rev2}
Compared to the existing studies on the parameter inference of coupled oscillator models, our study is novel and distinguishable in the following three aspects. First, our approach relies solely on time series of the macroscopic order parameter, in contrast to the previous studies that require data from individual oscillators \cite{smelyanskiy2005reconstruction, stankovski2015coupling, stankovski2017coupling, tokuda2019practical, matsuki2024network, su2025pairwise}. In this sense, our method addresses the inference problem under a more realistic and practically relevant setting. We further utilize an analytical solution of the Kuramoto order parameter during the relaxation process, derived in the thermodynamic limit of the system size. This represents an alternative strategy to the approach in Ref.~\cite{yamaguchi2024reconstruction}, which is based on linear response theory and requires weak perturbations to the system. Incorporating the analytical solution into the estimation process not only reduces computational cost but also presents a valuable integration of theoretical results from statistical physics with parameter inference problems. 


Second, we adopt a Bayesian approach to evaluate estimation performance by comparing both the estimated parameters and the corresponding posterior distributions. In particular, by examining the positional relationship among the true parameter values, the MAP estimators, and the posterior distributions, we clarify a fundamental difference between the benchmark and the model-mismatch cases, as illustrated in the contrast between Figs.~\ref{fig:OA_rep} (g), (h) and Figs.~\ref{fig:kura_rep} (g), (h). These findings would not be captured using conventional inference techniques, such as least-squares regression or maximum likelihood estimation, which yield only point estimates. 

Lastly, we analyze the long-term behavior of the Kuramoto order parameter and calculate the autocorrelation of its deviation from the analytical solution. As shown in Fig.~\ref{fig:long_data}, this deviation can be well approximated by an OU process, consistent with recent findings in Ref.~\cite{yue2024stochastic}. Based on this observation, we propose a modified estimation model [Eq.~\eqref{eq:data2}] that may lead to improved estimation accuracy. The stochastic model reduction of complex dynamical systems has become a topic of growing interest in recent years~\cite{snyder2021data, yue2024stochastic}. Our framework provides a possible approach for applying such reduced models, such as the OU model-based formulation in Eq.~\eqref{eq:data2}, to parameter inference.

In conclusion, we estimate the parameters of the Kuramoto model from macroscopic observations by leveraging theoretical results from the Ott–Antonsen ansatz. For efficient estimation of the posterior distribution and marginal likelihood, we adopt the EMC method. To the best of our knowledge, this is the first application of the EMC method in the context of nonlinear oscillator systems. After validating the internal consistency of our framework using a benchmark case, we apply it to datasets generated from the Kuramoto model. The inferred parameters converge to the true values as the system size $N$ increases. However, the positional relationship between the posterior distributions and the true parameter values suggests that the white Gaussian noise assumption in the estimation model [Eq.~\eqref{eq:data}] may require refinement. Incorporating the OU process-based estimation model [Eq.~\eqref{eq:data2}] into our Bayesian framework, as well as expanding our framework to enable model comparison, remain significant future challenges.

\section*{Data availability}
The data and simulation codes used in the present article are available in the following GitHub repository: \url{https://github.com/yuu-kato/EMC_kuramoto_2025}.

\begin{acknowledgments}
Y.K. thanks H. Ishii and J. Albrecht for valuable discussion, especially for the technical advice on the EMC simulation codes.
Y.K. also thanks R. Gro{\ss}mann, C. Beta,  O. Omelchenko, M. Rosenblum, T. Carletti, and M. Moriam\'e for fruitful discussion. 
This study was supported by 
the JSPS KAKENHI (No.\ JP23H00486) to M.O., 
the JSPS KAKENHI (No.\ JP23KJ0756) to Y.K., 
the JSPS KAKENHI (No.\ JP23KJ0723) to S.K., 
the JSPS KAKENHI (No.\ JP21J23250) to E.W., 
and the Graduate School of Frontier Sciences, The University of Tokyo, through the Challenging New Area Doctoral Research Grant (Project No.\ C2407) to Y.K. and S.K.
\end{acknowledgments}

\section*{Author contributions}
Y.K., S.K., E.W., M.O., and H.K. conceived this project and discussed the research direction. S.K. provided the original codes for the EMC estimation and its visualization. Y.K. revised the simulation codes with support from S.K. Y.K. mathematically formulated the theoretical and estimation framework, conducted the numerical experiments, analyzed the results, produced all the figures in this article, and wrote the manuscript with input from the rest of the authors. All authors confirmed the final version of the manuscript. M.O. and H.K. supervised the project.

\appendix
\section{Optimization of inverse temperature by hierarchical Bayes approach}
\label{sec:app_hier}
In the hierarchical Bayes method, the hyperparameter $b$ is regarded as a random variable with a prior distribution $p(b)$. Then, the joint posterior probability $p(\theta, b|D)$ is expressed as 
\begin{align}
    p(\theta, b| D) &= \frac{p(D, \theta, b)}{\displaystyle \int d b \int d \theta p(D, \theta, b)} \notag \\
    &= \frac{p(D| \theta, b) p(\theta) p(b)}{\displaystyle \int d b \int d \theta p(D| \theta, b) p(\theta) p(b)} \notag 
\end{align}
By marginalizing over $\theta$, we can calculate the posterior probability over the discrete set of inverse temperatures $\{b_l\}$ as follows:
\begin{align}
    p(b_l | D) 
    &= \frac{\displaystyle\int d\theta p(D | \theta, b_l) p(\theta) p(b_l)}{\displaystyle\sum_{l} \displaystyle\int d\theta p(D | \theta, b_l) p(\theta) p(b_l)} \notag \\
    &= \frac{Z(b_l) p(b_l)}{\displaystyle\sum_{l} Z(b_l) p(b_l)} \notag \\
    &= \frac{Z(b_l)}{\displaystyle\sum_{l} Z(b_l)}. \label{eq:pos_bl}
\end{align}

In the last equation, we assume that the prior density p($b_l$) is a discrete uniform distribution, i.e., $p(b_l) = 1/L$. Note that, in this case, the value of $b_l$ that maximizes the posterior density in Eq.~\eqref{eq:pos_bl} coincides with the value $b_{\hat{l}}$ obtained by maximizing the marginal likelihood in Eq.\ \eqref{eq:inv_opt}. 

\section{Convergence of Estimation Error in Monte Carlo Simulation}
\label{sec:app_err}
\begin{figure}
    \includegraphics[width=.98\linewidth]{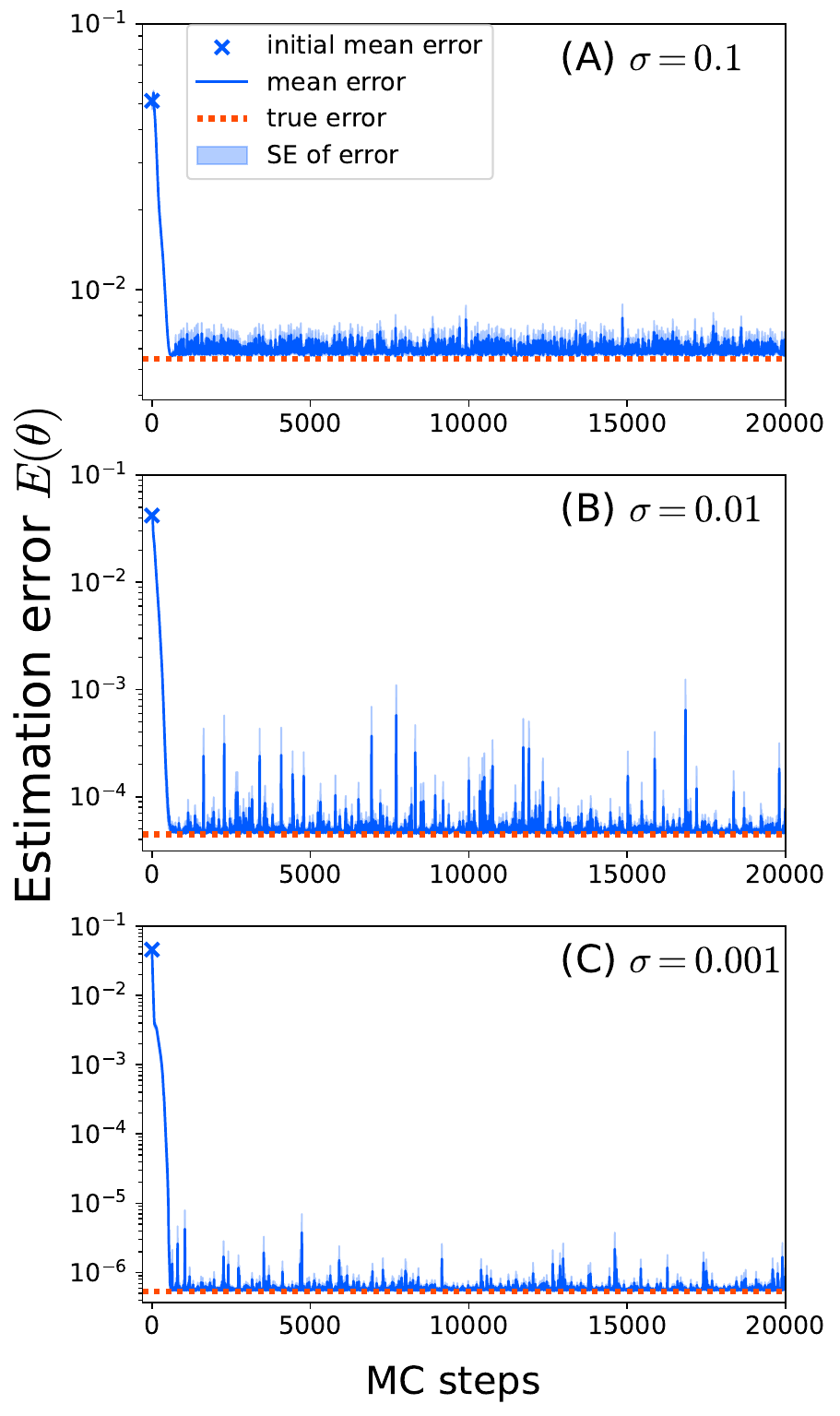}%
    \caption{The estimation error $E(\theta)$, given by Eq.~\eqref{eq:error}, is shown as a function of MC steps. Panels (A), (B), and (C) correspond to the artificial datasets presented in Figs.~\ref{fig:OA_single} (A-1), (B-1), and (C-1), respectively. For each dataset, 100 individual EMC simulations are performed. The mean estimation error is plotted as the blue line, and the standard error (SE) is shown by the blue shaded region, both evaluated at every 10 MC steps. 
    To highlight convergence behavior, the initial error and true error are shown with a cross mark and the red dotted line, respectively.}
    \label{fig:OA_error}
\end{figure}

\begin{figure}
    \includegraphics[width=.98\linewidth]{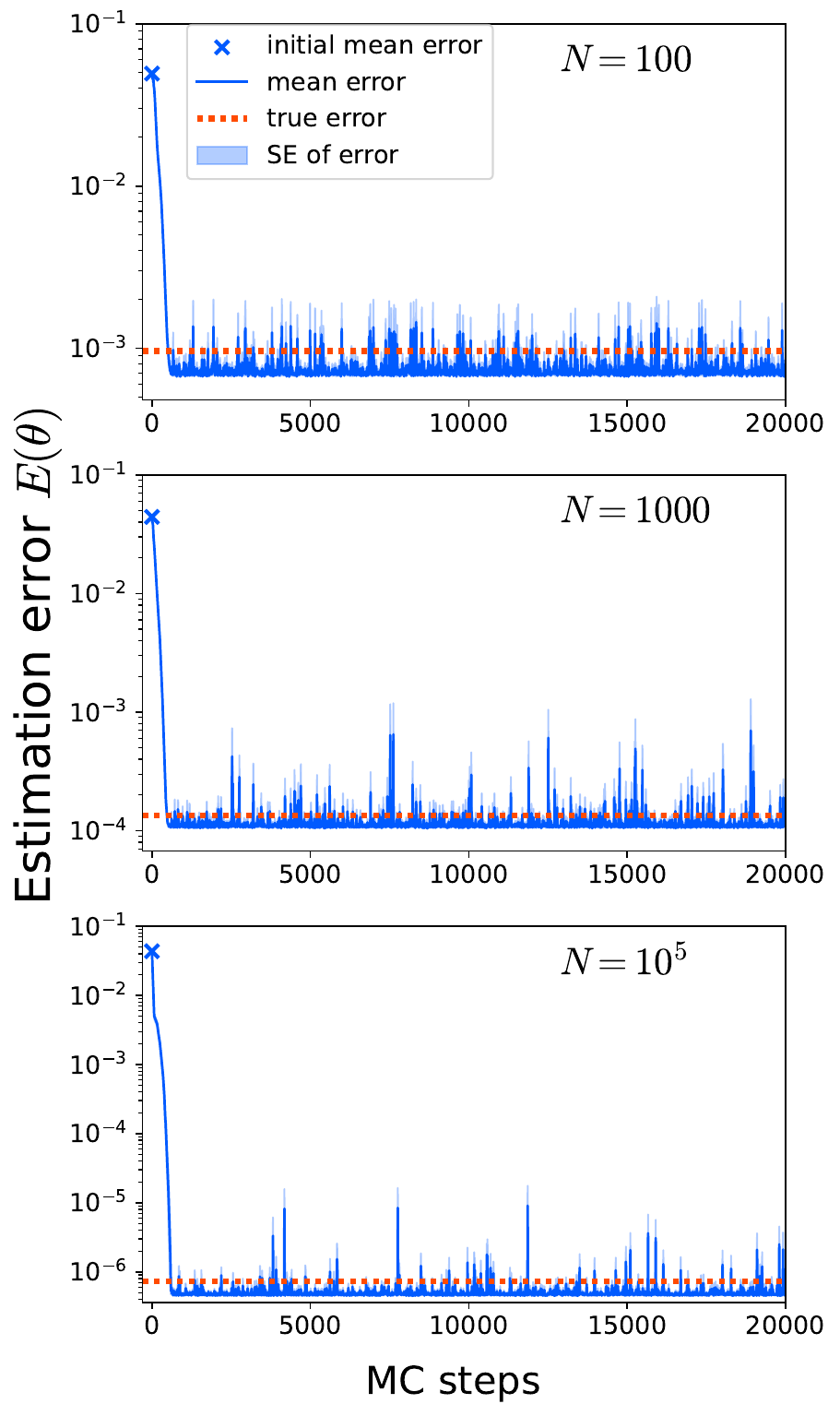}%
    \caption{The estimation error $E(\theta)$ as a function of MC steps. Panels (A), (B), and (C) correspond to the artificial datasets presented in Figs.~\ref{fig:kura_single} (A-1), (B-1), and (C-1), respectively.
    The error function is recorded, analyzed, and visualized in the same way as in Fig.~\ref{fig:OA_error}. }
    \label{fig:kura_error}
\end{figure}

To evaluate the convergence of the EMC simulations in Figs.~\ref{fig:OA_single} and \ref{fig:kura_single}, we plot the estimation error $E(\theta)$ as a function of MC steps in Figs.~\ref{fig:OA_error} and \ref{fig:kura_error}. In all panels, the estimation error approaches the true error, which is calculated by substituting the true parameter ($K=0.05,\ \gamma=0.08$) into Eq.~\eqref{eq:error}.
The convergence is achieved within 20000 MC steps, indicating that the choice of $S=100000$ (total number of MC steps) is sufficient to ensure reliable convergence of the simulation. It is noteworthy that, in the case where datasets are generated from the Kuramoto model (Fig.~\ref{fig:kura_error}), the estimation error converges to a value slightly lower than the true error. We consider that this discrepancy reflects the model mismatch between the data generation [Eq.~\eqref{eq:kura}] and the estimation process [Eq.~\eqref{eq:data}]. 

\section{Underlying cause of the parameter correlation in the 2D histogram}
\label{sec:app_corr}
\begin{figure}
    \includegraphics[width=.98\linewidth]{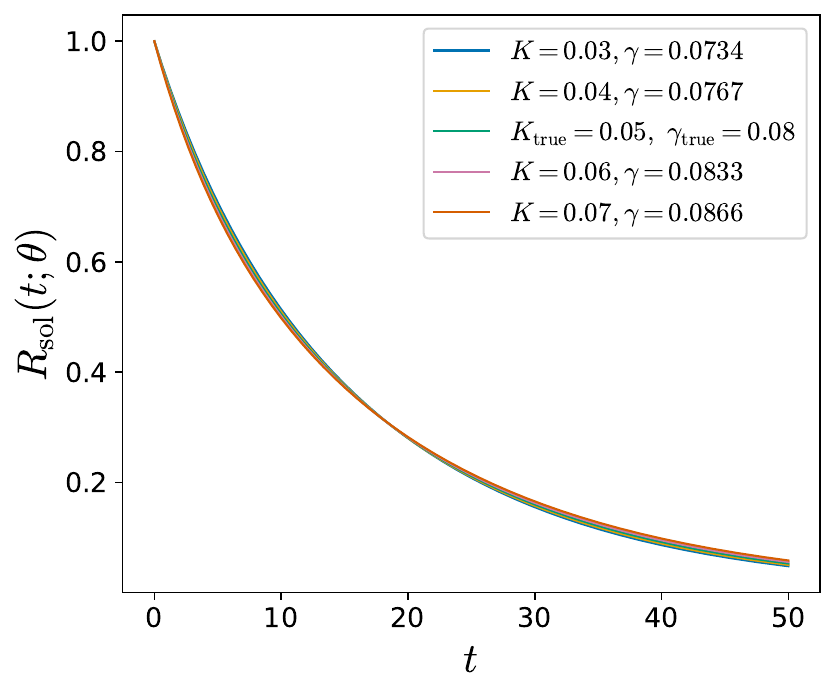}%
    \caption{Shapes of the analytical solution $R_{\rm sol}(t;\theta)$ [Eq.~\eqref{eq:oa_sol}] for different combinations of $K$ and $\gamma$ sampled along the regression line~\eqref{eq:corr}. The five colored curves exhibit nearly identical shapes. 
    }
    \label{fig:correlations}
\end{figure}

In this section, we aim to elucidate the origin of the linear correlation [Eq.~\eqref{eq:corr}] between $K$ and $\gamma$, as observed in the two-dimensional histograms in Figs.~\ref{fig:OA_single} and \ref{fig:kura_single}. Figure~\ref{fig:correlations} displays the shapes of analytical solutions~\eqref{eq:oa_sol} for various combinations of $K$ and $\gamma$ sampled along the regression line~\eqref{eq:corr}. We find that all five resulting curves exhibit nearly identical shapes. This observation suggests that the direction of the regression line corresponds to a trajectory in the $(K, \gamma)$ parameter space along which the functional form of $R_{\rm sol}(t;\theta)$ is approximately preserved. We therefore hypothesize that this direction corresponds to a trajectory in parameter space that minimizes the variation in the functional form of the analytical solution, thereby reducing the change in the estimation error [Eq.~\eqref{eq:error}] when fitting to the given datasets.


To confirm the above hypothesis, we formulate the following minimization problem.
We treat the analytical solution $R_{\rm sol}(t;\theta)$ in Eq.~\eqref{eq:oa_sol} as a function of $K$ and $\gamma$, denoted by $R_{\rm sol}(t;K, \gamma)$. Let $\Delta r \cos \theta$ and $\Delta r \sin \theta$ be small perturbations in $K$ and $\gamma$, where $\Delta r, \theta \in \R$ and $\Delta r$ is assumed to be sufficiently small. Our goal is to identify the angle $\theta$ that minimizes the overall change in the analytical solution:
\begin{align}
    \Delta \tilde R &\coloneqq \int_{0}^{50} \biggl[R_{\rm sol}(t;K_{\rm true} + \Delta r \cos \theta, \gamma_{\rm true} + \Delta r \sin \theta) \notag \\ 
    & \qquad - R_{\rm sol}(t;K_{\rm true}, \gamma_{\rm true})\biggr]^2 dt. \label{eq:minimize}
\end{align}
Expanding $R_{\rm sol}(t;K, \gamma)$ in a Taylor series around $(K_{\rm true}, \gamma_{\rm true})$ and neglecting higher-order terms in $\Delta r$, we have
\begin{align}
    \label{eq:minimize_original}
    \Delta \tilde R = \Delta r^2 (\zeta \cos^2 \theta + 2 \eta \cos \theta \sin \theta + \kappa \sin^2 \theta),
\end{align}
where the coefficients are given by
\begin{align}
    \zeta &\coloneqq \int_{0}^{50} \biggl[\frac{\del R_{\rm sol}}{\del K}(t;K_{\rm true}, \gamma_{\rm true}) \biggr]^2 dt, \label{eq:def_zeta}\\
    \eta &\coloneqq \int_{0}^{50} \biggl[\frac{\del R_{\rm sol}}{\del K}(t;K_{\rm true}, \gamma_{\rm true}) \biggr] \notag \\
    & \qquad \qquad \times \biggl[\frac{\del R_{\rm sol}}{\del \gamma}(t;K_{\rm true}, \gamma_{\rm true}) \biggr] dt, \label{eq:def_eta} \\
    \kappa &\coloneqq \int_{0}^{50} \biggl[\frac{\del R_{\rm sol}}{\del \gamma}(t;K_{\rm true}, \gamma_{\rm true}) \biggr]^2 dt. \label{eq:def_kappa}
\end{align}
Equation~\eqref{eq:minimize_original} can be rewritten as
\begin{align}
    \Delta \tilde R &= \Delta r^2 \left\{ \frac{\zeta + \kappa}{2} + \frac{\zeta - \kappa}{2} \cos 2\theta + \eta \sin 2\theta \right\} \notag \\
    &= \Delta r^2 \left\{ \frac{\zeta + \kappa}{2} + \alpha \sin (2\theta + \beta) \right\}, \label{eq:minimize_final}
\end{align}
where
\begin{align}
    \alpha &\coloneqq \sqrt{\frac{(\zeta - \kappa)^2}{4} + \eta^2}, 
\end{align}
and $\beta$ is set by
\begin{gather}
    \cos \beta = \frac{\eta}{\alpha}, \quad
    \sin \beta = \frac{\zeta - \kappa}{2 \alpha}. 
\end{gather}
According to Eq.~\eqref{eq:minimize_final}, the optimal $\theta$ that minimizes $\Delta \tilde R$ is given by 
\begin{equation}
    \theta = \theta^* \coloneqq \frac{3\pi}{4} - \frac{\beta}{2}.
\end{equation}
Thus, the optimal slope in the $(K, \gamma)$ parameter space, which minimizes changes in the analytical solution, is 
\begin{align}
    \tan \theta^* &= \frac{-1-\tan \frac{\beta}{2}}{1-\tan \frac{\beta}{2}} \notag \\
    &= \frac{-1 -\cos \beta - \sin \beta}{1 +\cos \beta - \sin \beta} \notag \\
    &= \frac{-\sqrt{\frac{(\zeta - \kappa)^2}{4} + \eta^2} - \eta - \frac{\zeta - \kappa}{2}}{\sqrt{\frac{(\zeta - \kappa)^2}{4} + \eta^2} + \eta -\frac{\zeta - \kappa}{2}},
\end{align}
where we use the formula $\tan \frac{\beta}{2} = \frac{\sin \beta}{1+\cos \beta}$. By numerically evaluating Eqs.~\eqref{eq:def_zeta}--\eqref{eq:def_kappa}, we finally obtain
\begin{equation}
    \label{eq:slope_num}
    \tan \theta^* \simeq 0.339, 
\end{equation}
which is in close agreement with the slope of the empirical regression line in Eq.~\eqref{eq:corr}. The numerical result in Eq.~\eqref{eq:slope_num} is obtained using {\it Mathematica}; the corresponding code is available in our GitHub repository as described in Data Availability section.  

\bibliography{papers.bib}

\begin{thebibliography}{45}%
\makeatletter
\providecommand \@ifxundefined [1]{%
 \@ifx{#1\undefined}
}%
\providecommand \@ifnum [1]{%
 \ifnum #1\expandafter \@firstoftwo
 \else \expandafter \@secondoftwo
 \fi
}%
\providecommand \@ifx [1]{%
 \ifx #1\expandafter \@firstoftwo
 \else \expandafter \@secondoftwo
 \fi
}%
\providecommand \natexlab [1]{#1}%
\providecommand \enquote  [1]{``#1''}%
\providecommand \bibnamefont  [1]{#1}%
\providecommand \bibfnamefont [1]{#1}%
\providecommand \citenamefont [1]{#1}%
\providecommand \href@noop [0]{\@secondoftwo}%
\providecommand \href [0]{\begingroup \@sanitize@url \@href}%
\providecommand \@href[1]{\@@startlink{#1}\@@href}%
\providecommand \@@href[1]{\endgroup#1\@@endlink}%
\providecommand \@sanitize@url [0]{\catcode `\\12\catcode `\$12\catcode `\&12\catcode `\#12\catcode `\^12\catcode `\_12\catcode `\%12\relax}%
\providecommand \@@startlink[1]{}%
\providecommand \@@endlink[0]{}%
\providecommand \url  [0]{\begingroup\@sanitize@url \@url }%
\providecommand \@url [1]{\endgroup\@href {#1}{\urlprefix }}%
\providecommand \urlprefix  [0]{URL }%
\providecommand \Eprint [0]{\href }%
\providecommand \doibase [0]{https://doi.org/}%
\providecommand \selectlanguage [0]{\@gobble}%
\providecommand \bibinfo  [0]{\@secondoftwo}%
\providecommand \bibfield  [0]{\@secondoftwo}%
\providecommand \translation [1]{[#1]}%
\providecommand \BibitemOpen [0]{}%
\providecommand \bibitemStop [0]{}%
\providecommand \bibitemNoStop [0]{.\EOS\space}%
\providecommand \EOS [0]{\spacefactor3000\relax}%
\providecommand \BibitemShut  [1]{\csname bibitem#1\endcsname}%
\let\auto@bib@innerbib\@empty
\bibitem [{\citenamefont {Pikovsky}\ \emph {et~al.}(2002)\citenamefont {Pikovsky}, \citenamefont {Rosenblum},\ and\ \citenamefont {Kurths}}]{pikovsky2002synchronization}%
  \BibitemOpen
  \bibfield  {author} {\bibinfo {author} {\bibfnamefont {A.}~\bibnamefont {Pikovsky}}, \bibinfo {author} {\bibfnamefont {M.}~\bibnamefont {Rosenblum}},\ and\ \bibinfo {author} {\bibfnamefont {J.}~\bibnamefont {Kurths}},\ }\href@noop {} {\emph {\bibinfo {title} {Synchronization: a universal concept in nonlinear science}}}\ (\bibinfo  {publisher} {Cambridge University Press},\ \bibinfo {year} {2002})\BibitemShut {NoStop}%
\bibitem [{\citenamefont {Smith}(1935)}]{smith1935synchronous}%
  \BibitemOpen
  \bibfield  {author} {\bibinfo {author} {\bibfnamefont {H.~M.}\ \bibnamefont {Smith}},\ }\bibfield  {title} {\bibinfo {title} {Synchronous flashing of fireflies},\ }\href@noop {} {\bibfield  {journal} {\bibinfo  {journal} {Science}\ }\textbf {\bibinfo {volume} {82}},\ \bibinfo {pages} {151} (\bibinfo {year} {1935})}\BibitemShut {NoStop}%
\bibitem [{\citenamefont {Uhlhaas}\ and\ \citenamefont {Singer}(2006)}]{uhlhaas2006neural}%
  \BibitemOpen
  \bibfield  {author} {\bibinfo {author} {\bibfnamefont {P.~J.}\ \bibnamefont {Uhlhaas}}\ and\ \bibinfo {author} {\bibfnamefont {W.}~\bibnamefont {Singer}},\ }\bibfield  {title} {\bibinfo {title} {Neural synchrony in brain disorders: relevance for cognitive dysfunctions and pathophysiology},\ }\href@noop {} {\bibfield  {journal} {\bibinfo  {journal} {Neuron}\ }\textbf {\bibinfo {volume} {52}},\ \bibinfo {pages} {155} (\bibinfo {year} {2006})}\BibitemShut {NoStop}%
\bibitem [{\citenamefont {Hammond}\ \emph {et~al.}(2007)\citenamefont {Hammond}, \citenamefont {Bergman},\ and\ \citenamefont {Brown}}]{hammond2007pathological}%
  \BibitemOpen
  \bibfield  {author} {\bibinfo {author} {\bibfnamefont {C.}~\bibnamefont {Hammond}}, \bibinfo {author} {\bibfnamefont {H.}~\bibnamefont {Bergman}},\ and\ \bibinfo {author} {\bibfnamefont {P.}~\bibnamefont {Brown}},\ }\bibfield  {title} {\bibinfo {title} {Pathological synchronization in {P}arkinson's disease: networks, models and treatments},\ }\href@noop {} {\bibfield  {journal} {\bibinfo  {journal} {Trends in Neurosciences}\ }\textbf {\bibinfo {volume} {30}},\ \bibinfo {pages} {357} (\bibinfo {year} {2007})}\BibitemShut {NoStop}%
\bibitem [{\citenamefont {Jiruska}\ \emph {et~al.}(2013)\citenamefont {Jiruska}, \citenamefont {De~Curtis}, \citenamefont {Jefferys}, \citenamefont {Schevon}, \citenamefont {Schiff},\ and\ \citenamefont {Schindler}}]{jiruska2013synchronization}%
  \BibitemOpen
  \bibfield  {author} {\bibinfo {author} {\bibfnamefont {P.}~\bibnamefont {Jiruska}}, \bibinfo {author} {\bibfnamefont {M.}~\bibnamefont {De~Curtis}}, \bibinfo {author} {\bibfnamefont {J.~G.}\ \bibnamefont {Jefferys}}, \bibinfo {author} {\bibfnamefont {C.~A.}\ \bibnamefont {Schevon}}, \bibinfo {author} {\bibfnamefont {S.~J.}\ \bibnamefont {Schiff}},\ and\ \bibinfo {author} {\bibfnamefont {K.}~\bibnamefont {Schindler}},\ }\bibfield  {title} {\bibinfo {title} {Synchronization and desynchronization in epilepsy: controversies and hypotheses},\ }\href@noop {} {\bibfield  {journal} {\bibinfo  {journal} {The Journal of Physiology}\ }\textbf {\bibinfo {volume} {591}},\ \bibinfo {pages} {787} (\bibinfo {year} {2013})}\BibitemShut {NoStop}%
\bibitem [{\citenamefont {Strogatz}\ \emph {et~al.}(2005)\citenamefont {Strogatz}, \citenamefont {Abrams}, \citenamefont {McRobie}, \citenamefont {Eckhardt},\ and\ \citenamefont {Ott}}]{strogatz2005crowd}%
  \BibitemOpen
  \bibfield  {author} {\bibinfo {author} {\bibfnamefont {S.~H.}\ \bibnamefont {Strogatz}}, \bibinfo {author} {\bibfnamefont {D.~M.}\ \bibnamefont {Abrams}}, \bibinfo {author} {\bibfnamefont {A.}~\bibnamefont {McRobie}}, \bibinfo {author} {\bibfnamefont {B.}~\bibnamefont {Eckhardt}},\ and\ \bibinfo {author} {\bibfnamefont {E.}~\bibnamefont {Ott}},\ }\bibfield  {title} {\bibinfo {title} {Crowd synchrony on the millennium bridge},\ }\href@noop {} {\bibfield  {journal} {\bibinfo  {journal} {Nature}\ }\textbf {\bibinfo {volume} {438}},\ \bibinfo {pages} {43} (\bibinfo {year} {2005})}\BibitemShut {NoStop}%
\bibitem [{\citenamefont {Kuramoto}(1984)}]{kuramoto84}%
  \BibitemOpen
  \bibfield  {author} {\bibinfo {author} {\bibfnamefont {Y.}~\bibnamefont {Kuramoto}},\ }\href@noop {} {\emph {\bibinfo {title} {Chemical Oscillations, Waves, and Turbulence}}}\ (\bibinfo  {publisher} {Springer},\ \bibinfo {address} {New York},\ \bibinfo {year} {1984})\BibitemShut {NoStop}%
\bibitem [{\citenamefont {Acebr{\'o}n}\ \emph {et~al.}(2005)\citenamefont {Acebr{\'o}n}, \citenamefont {Bonilla}, \citenamefont {P{\'e}rez~Vicente}, \citenamefont {Ritort},\ and\ \citenamefont {Spigler}}]{acebron2005kuramoto}%
  \BibitemOpen
  \bibfield  {author} {\bibinfo {author} {\bibfnamefont {J.~A.}\ \bibnamefont {Acebr{\'o}n}}, \bibinfo {author} {\bibfnamefont {L.~L.}\ \bibnamefont {Bonilla}}, \bibinfo {author} {\bibfnamefont {C.~J.}\ \bibnamefont {P{\'e}rez~Vicente}}, \bibinfo {author} {\bibfnamefont {F.}~\bibnamefont {Ritort}},\ and\ \bibinfo {author} {\bibfnamefont {R.}~\bibnamefont {Spigler}},\ }\bibfield  {title} {\bibinfo {title} {The {K}uramoto model: A simple paradigm for synchronization phenomena},\ }\href@noop {} {\bibfield  {journal} {\bibinfo  {journal} {Reviews of Modern Physics}\ }\textbf {\bibinfo {volume} {77}},\ \bibinfo {pages} {137} (\bibinfo {year} {2005})}\BibitemShut {NoStop}%
\bibitem [{\citenamefont {Rodrigues}\ \emph {et~al.}(2016)\citenamefont {Rodrigues}, \citenamefont {Peron}, \citenamefont {Ji},\ and\ \citenamefont {Kurths}}]{rodrigues2016kuramoto}%
  \BibitemOpen
  \bibfield  {author} {\bibinfo {author} {\bibfnamefont {F.~A.}\ \bibnamefont {Rodrigues}}, \bibinfo {author} {\bibfnamefont {T.~K.~D.}\ \bibnamefont {Peron}}, \bibinfo {author} {\bibfnamefont {P.}~\bibnamefont {Ji}},\ and\ \bibinfo {author} {\bibfnamefont {J.}~\bibnamefont {Kurths}},\ }\bibfield  {title} {\bibinfo {title} {The {K}uramoto model in complex networks},\ }\href@noop {} {\bibfield  {journal} {\bibinfo  {journal} {Physics Reports}\ }\textbf {\bibinfo {volume} {610}},\ \bibinfo {pages} {1} (\bibinfo {year} {2016})}\BibitemShut {NoStop}%
\bibitem [{\citenamefont {Pietras}\ and\ \citenamefont {Daffertshofer}(2019)}]{pietras2019network}%
  \BibitemOpen
  \bibfield  {author} {\bibinfo {author} {\bibfnamefont {B.}~\bibnamefont {Pietras}}\ and\ \bibinfo {author} {\bibfnamefont {A.}~\bibnamefont {Daffertshofer}},\ }\bibfield  {title} {\bibinfo {title} {Network dynamics of coupled oscillators and phase reduction techniques},\ }\href@noop {} {\bibfield  {journal} {\bibinfo  {journal} {Physics Reports}\ }\textbf {\bibinfo {volume} {819}},\ \bibinfo {pages} {1} (\bibinfo {year} {2019})}\BibitemShut {NoStop}%
\bibitem [{\citenamefont {Cumin}\ and\ \citenamefont {Unsworth}(2007)}]{cumin2007generalising}%
  \BibitemOpen
  \bibfield  {author} {\bibinfo {author} {\bibfnamefont {D.}~\bibnamefont {Cumin}}\ and\ \bibinfo {author} {\bibfnamefont {C.}~\bibnamefont {Unsworth}},\ }\bibfield  {title} {\bibinfo {title} {Generalising the {K}uramoto model for the study of neuronal synchronisation in the brain},\ }\href@noop {} {\bibfield  {journal} {\bibinfo  {journal} {Physica D: Nonlinear Phenomena}\ }\textbf {\bibinfo {volume} {226}},\ \bibinfo {pages} {181} (\bibinfo {year} {2007})}\BibitemShut {NoStop}%
\bibitem [{\citenamefont {Filatrella}\ \emph {et~al.}(2008)\citenamefont {Filatrella}, \citenamefont {Nielsen},\ and\ \citenamefont {Pedersen}}]{filatrella2008analysis}%
  \BibitemOpen
  \bibfield  {author} {\bibinfo {author} {\bibfnamefont {G.}~\bibnamefont {Filatrella}}, \bibinfo {author} {\bibfnamefont {A.~H.}\ \bibnamefont {Nielsen}},\ and\ \bibinfo {author} {\bibfnamefont {N.~F.}\ \bibnamefont {Pedersen}},\ }\bibfield  {title} {\bibinfo {title} {Analysis of a power grid using a {K}uramoto-like model},\ }\href@noop {} {\bibfield  {journal} {\bibinfo  {journal} {The European Physical Journal B}\ }\textbf {\bibinfo {volume} {61}},\ \bibinfo {pages} {485} (\bibinfo {year} {2008})}\BibitemShut {NoStop}%
\bibitem [{\citenamefont {Potratzki}\ \emph {et~al.}(2024)\citenamefont {Potratzki}, \citenamefont {Br{\"o}hl}, \citenamefont {Rings},\ and\ \citenamefont {Lehnertz}}]{potratzki2024synchronization}%
  \BibitemOpen
  \bibfield  {author} {\bibinfo {author} {\bibfnamefont {M.}~\bibnamefont {Potratzki}}, \bibinfo {author} {\bibfnamefont {T.}~\bibnamefont {Br{\"o}hl}}, \bibinfo {author} {\bibfnamefont {T.}~\bibnamefont {Rings}},\ and\ \bibinfo {author} {\bibfnamefont {K.}~\bibnamefont {Lehnertz}},\ }\bibfield  {title} {\bibinfo {title} {Synchronization dynamics of phase oscillators on power grid models},\ }\href@noop {} {\bibfield  {journal} {\bibinfo  {journal} {Chaos: An Interdisciplinary Journal of Nonlinear Science}\ }\textbf {\bibinfo {volume} {34}} (\bibinfo {year} {2024})}\BibitemShut {NoStop}%
\bibitem [{\citenamefont {Wang}\ and\ \citenamefont {Roychowdhury}(2017)}]{wang2017oscillator}%
  \BibitemOpen
  \bibfield  {author} {\bibinfo {author} {\bibfnamefont {T.}~\bibnamefont {Wang}}\ and\ \bibinfo {author} {\bibfnamefont {J.}~\bibnamefont {Roychowdhury}},\ }\bibfield  {title} {\bibinfo {title} {Oscillator-based ising machine},\ }\href@noop {} {\bibfield  {journal} {\bibinfo  {journal} {arXiv preprint arXiv:1709.08102}\ } (\bibinfo {year} {2017})}\BibitemShut {NoStop}%
\bibitem [{\citenamefont {Smelyanskiy}\ \emph {et~al.}(2005)\citenamefont {Smelyanskiy}, \citenamefont {Luchinsky}, \citenamefont {Timucin},\ and\ \citenamefont {Bandrivskyy}}]{smelyanskiy2005reconstruction}%
  \BibitemOpen
  \bibfield  {author} {\bibinfo {author} {\bibfnamefont {V.}~\bibnamefont {Smelyanskiy}}, \bibinfo {author} {\bibfnamefont {D.~G.}\ \bibnamefont {Luchinsky}}, \bibinfo {author} {\bibfnamefont {D.}~\bibnamefont {Timucin}},\ and\ \bibinfo {author} {\bibfnamefont {A.}~\bibnamefont {Bandrivskyy}},\ }\bibfield  {title} {\bibinfo {title} {Reconstruction of stochastic nonlinear dynamical models from trajectory measurements},\ }\href@noop {} {\bibfield  {journal} {\bibinfo  {journal} {Physical Review E}\ }\textbf {\bibinfo {volume} {72}},\ \bibinfo {pages} {026202} (\bibinfo {year} {2005})}\BibitemShut {NoStop}%
\bibitem [{\citenamefont {Stankovski}\ \emph {et~al.}(2015)\citenamefont {Stankovski}, \citenamefont {Ticcinelli}, \citenamefont {McClintock},\ and\ \citenamefont {Stefanovska}}]{stankovski2015coupling}%
  \BibitemOpen
  \bibfield  {author} {\bibinfo {author} {\bibfnamefont {T.}~\bibnamefont {Stankovski}}, \bibinfo {author} {\bibfnamefont {V.}~\bibnamefont {Ticcinelli}}, \bibinfo {author} {\bibfnamefont {P.~V.}\ \bibnamefont {McClintock}},\ and\ \bibinfo {author} {\bibfnamefont {A.}~\bibnamefont {Stefanovska}},\ }\bibfield  {title} {\bibinfo {title} {Coupling functions in networks of oscillators},\ }\href@noop {} {\bibfield  {journal} {\bibinfo  {journal} {New Journal of Physics}\ }\textbf {\bibinfo {volume} {17}},\ \bibinfo {pages} {035002} (\bibinfo {year} {2015})}\BibitemShut {NoStop}%
\bibitem [{\citenamefont {Stankovski}\ \emph {et~al.}(2017)\citenamefont {Stankovski}, \citenamefont {Pereira}, \citenamefont {McClintock},\ and\ \citenamefont {Stefanovska}}]{stankovski2017coupling}%
  \BibitemOpen
  \bibfield  {author} {\bibinfo {author} {\bibfnamefont {T.}~\bibnamefont {Stankovski}}, \bibinfo {author} {\bibfnamefont {T.}~\bibnamefont {Pereira}}, \bibinfo {author} {\bibfnamefont {P.~V.}\ \bibnamefont {McClintock}},\ and\ \bibinfo {author} {\bibfnamefont {A.}~\bibnamefont {Stefanovska}},\ }\bibfield  {title} {\bibinfo {title} {Coupling functions: universal insights into dynamical interaction mechanisms},\ }\href@noop {} {\bibfield  {journal} {\bibinfo  {journal} {Reviews of Modern Physics}\ }\textbf {\bibinfo {volume} {89}},\ \bibinfo {pages} {045001} (\bibinfo {year} {2017})}\BibitemShut {NoStop}%
\bibitem [{\citenamefont {Tokuda}\ \emph {et~al.}(2019)\citenamefont {Tokuda}, \citenamefont {Levnajic},\ and\ \citenamefont {Ishimura}}]{tokuda2019practical}%
  \BibitemOpen
  \bibfield  {author} {\bibinfo {author} {\bibfnamefont {I.~T.}\ \bibnamefont {Tokuda}}, \bibinfo {author} {\bibfnamefont {Z.}~\bibnamefont {Levnajic}},\ and\ \bibinfo {author} {\bibfnamefont {K.}~\bibnamefont {Ishimura}},\ }\bibfield  {title} {\bibinfo {title} {A practical method for estimating coupling functions in complex dynamical systems},\ }\href@noop {} {\bibfield  {journal} {\bibinfo  {journal} {Philosophical Transactions of the Royal Society A}\ }\textbf {\bibinfo {volume} {377}},\ \bibinfo {pages} {20190015} (\bibinfo {year} {2019})}\BibitemShut {NoStop}%
\bibitem [{\citenamefont {Matsuki}\ \emph {et~al.}(2024)\citenamefont {Matsuki}, \citenamefont {Kori},\ and\ \citenamefont {Kobayashi}}]{matsuki2024network}%
  \BibitemOpen
  \bibfield  {author} {\bibinfo {author} {\bibfnamefont {A.}~\bibnamefont {Matsuki}}, \bibinfo {author} {\bibfnamefont {H.}~\bibnamefont {Kori}},\ and\ \bibinfo {author} {\bibfnamefont {R.}~\bibnamefont {Kobayashi}},\ }\bibfield  {title} {\bibinfo {title} {Network inference from oscillatory signals based on circle map},\ }\href@noop {} {\bibfield  {journal} {\bibinfo  {journal} {arXiv preprint arXiv:2407.07445}\ } (\bibinfo {year} {2024})}\BibitemShut {NoStop}%
\bibitem [{\citenamefont {Su}\ \emph {et~al.}(2025)\citenamefont {Su}, \citenamefont {Hata}, \citenamefont {Kori}, \citenamefont {Nakao},\ and\ \citenamefont {Kobayashi}}]{su2025pairwise}%
  \BibitemOpen
  \bibfield  {author} {\bibinfo {author} {\bibfnamefont {W.}~\bibnamefont {Su}}, \bibinfo {author} {\bibfnamefont {S.}~\bibnamefont {Hata}}, \bibinfo {author} {\bibfnamefont {H.}~\bibnamefont {Kori}}, \bibinfo {author} {\bibfnamefont {H.}~\bibnamefont {Nakao}},\ and\ \bibinfo {author} {\bibfnamefont {R.}~\bibnamefont {Kobayashi}},\ }\bibfield  {title} {\bibinfo {title} {Pairwise vs higher-order interactions: Can we identify the interaction type in coupled oscillators from time series?},\ }\href@noop {} {\bibfield  {journal} {\bibinfo  {journal} {arXiv preprint arXiv:2503.13244}\ } (\bibinfo {year} {2025})}\BibitemShut {NoStop}%
\bibitem [{\citenamefont {Ozawa}\ and\ \citenamefont {Kori}(2024)}]{ozawa2024two}%
  \BibitemOpen
  \bibfield  {author} {\bibinfo {author} {\bibfnamefont {A.}~\bibnamefont {Ozawa}}\ and\ \bibinfo {author} {\bibfnamefont {H.}~\bibnamefont {Kori}},\ }\bibfield  {title} {\bibinfo {title} {Two distinct transitions in a population of coupled oscillators with turnover: desynchronization and stochastic oscillation quenching},\ }\href@noop {} {\bibfield  {journal} {\bibinfo  {journal} {Physical Review Letters}\ }\textbf {\bibinfo {volume} {133}},\ \bibinfo {pages} {047201} (\bibinfo {year} {2024})}\BibitemShut {NoStop}%
\bibitem [{\citenamefont {Ott}\ and\ \citenamefont {Antonsen}(2008)}]{ott2008low}%
  \BibitemOpen
  \bibfield  {author} {\bibinfo {author} {\bibfnamefont {E.}~\bibnamefont {Ott}}\ and\ \bibinfo {author} {\bibfnamefont {T.~M.}\ \bibnamefont {Antonsen}},\ }\bibfield  {title} {\bibinfo {title} {Low dimensional behavior of large systems of globally coupled oscillators},\ }\href@noop {} {\bibfield  {journal} {\bibinfo  {journal} {Chaos: An Interdisciplinary Journal of Nonlinear Science}\ }\textbf {\bibinfo {volume} {18}},\ \bibinfo {pages} {037113} (\bibinfo {year} {2008})}\BibitemShut {NoStop}%
\bibitem [{\citenamefont {Ott}\ and\ \citenamefont {Antonsen}(2009)}]{ott2009long}%
  \BibitemOpen
  \bibfield  {author} {\bibinfo {author} {\bibfnamefont {E.}~\bibnamefont {Ott}}\ and\ \bibinfo {author} {\bibfnamefont {T.~M.}\ \bibnamefont {Antonsen}},\ }\bibfield  {title} {\bibinfo {title} {Long time evolution of phase oscillator systems},\ }\href@noop {} {\bibfield  {journal} {\bibinfo  {journal} {Chaos: An interdisciplinary Journal of Nonlinear Science}\ }\textbf {\bibinfo {volume} {19}} (\bibinfo {year} {2009})}\BibitemShut {NoStop}%
\bibitem [{\citenamefont {Yamaguchi}\ and\ \citenamefont {Terada}(2024)}]{yamaguchi2024reconstruction}%
  \BibitemOpen
  \bibfield  {author} {\bibinfo {author} {\bibfnamefont {Y.~Y.}\ \bibnamefont {Yamaguchi}}\ and\ \bibinfo {author} {\bibfnamefont {Y.}~\bibnamefont {Terada}},\ }\bibfield  {title} {\bibinfo {title} {Reconstruction of phase dynamics from macroscopic observations based on linear and nonlinear response theories},\ }\href@noop {} {\bibfield  {journal} {\bibinfo  {journal} {Physical Review E}\ }\textbf {\bibinfo {volume} {109}},\ \bibinfo {pages} {024217} (\bibinfo {year} {2024})}\BibitemShut {NoStop}%
\bibitem [{\citenamefont {Daido}(1992)}]{daido1992order}%
  \BibitemOpen
  \bibfield  {author} {\bibinfo {author} {\bibfnamefont {H.}~\bibnamefont {Daido}},\ }\bibfield  {title} {\bibinfo {title} {Order function and macroscopic mutual entrainment in uniformly coupled limit-cycle oscillators},\ }\href@noop {} {\bibfield  {journal} {\bibinfo  {journal} {Progress of Theoretical Physics}\ }\textbf {\bibinfo {volume} {88}},\ \bibinfo {pages} {1213} (\bibinfo {year} {1992})}\BibitemShut {NoStop}%
\bibitem [{\citenamefont {Hukushima}\ and\ \citenamefont {Nemoto}(1996)}]{hukushima1996exchange}%
  \BibitemOpen
  \bibfield  {author} {\bibinfo {author} {\bibfnamefont {K.}~\bibnamefont {Hukushima}}\ and\ \bibinfo {author} {\bibfnamefont {K.}~\bibnamefont {Nemoto}},\ }\bibfield  {title} {\bibinfo {title} {Exchange {M}onte {C}arlo method and application to spin glass simulations},\ }\href@noop {} {\bibfield  {journal} {\bibinfo  {journal} {Journal of the Physical Society of Japan}\ }\textbf {\bibinfo {volume} {65}},\ \bibinfo {pages} {1604} (\bibinfo {year} {1996})}\BibitemShut {NoStop}%
\bibitem [{\citenamefont {Nagata}\ \emph {et~al.}(2012)\citenamefont {Nagata}, \citenamefont {Sugita},\ and\ \citenamefont {Okada}}]{nagata2012bayesian}%
  \BibitemOpen
  \bibfield  {author} {\bibinfo {author} {\bibfnamefont {K.}~\bibnamefont {Nagata}}, \bibinfo {author} {\bibfnamefont {S.}~\bibnamefont {Sugita}},\ and\ \bibinfo {author} {\bibfnamefont {M.}~\bibnamefont {Okada}},\ }\bibfield  {title} {\bibinfo {title} {Bayesian spectral deconvolution with the exchange {M}onte {C}arlo method},\ }\href@noop {} {\bibfield  {journal} {\bibinfo  {journal} {Neural Networks}\ }\textbf {\bibinfo {volume} {28}},\ \bibinfo {pages} {82} (\bibinfo {year} {2012})}\BibitemShut {NoStop}%
\bibitem [{\citenamefont {Tokuda}\ \emph {et~al.}(2017)\citenamefont {Tokuda}, \citenamefont {Nagata},\ and\ \citenamefont {Okada}}]{tokuda2017simultaneous}%
  \BibitemOpen
  \bibfield  {author} {\bibinfo {author} {\bibfnamefont {S.}~\bibnamefont {Tokuda}}, \bibinfo {author} {\bibfnamefont {K.}~\bibnamefont {Nagata}},\ and\ \bibinfo {author} {\bibfnamefont {M.}~\bibnamefont {Okada}},\ }\bibfield  {title} {\bibinfo {title} {Simultaneous estimation of noise variance and number of peaks in {B}ayesian spectral deconvolution},\ }\href@noop {} {\bibfield  {journal} {\bibinfo  {journal} {Journal of the Physical Society of Japan}\ }\textbf {\bibinfo {volume} {86}},\ \bibinfo {pages} {024001} (\bibinfo {year} {2017})}\BibitemShut {NoStop}%
\bibitem [{\citenamefont {Kashiwamura}\ \emph {et~al.}(2022)\citenamefont {Kashiwamura}, \citenamefont {Katakami}, \citenamefont {Yamagami}, \citenamefont {Iwamitsu}, \citenamefont {Kumazoe}, \citenamefont {Nagata}, \citenamefont {Okajima}, \citenamefont {Akai},\ and\ \citenamefont {Okada}}]{kashiwamura2022bayesian}%
  \BibitemOpen
  \bibfield  {author} {\bibinfo {author} {\bibfnamefont {S.}~\bibnamefont {Kashiwamura}}, \bibinfo {author} {\bibfnamefont {S.}~\bibnamefont {Katakami}}, \bibinfo {author} {\bibfnamefont {R.}~\bibnamefont {Yamagami}}, \bibinfo {author} {\bibfnamefont {K.}~\bibnamefont {Iwamitsu}}, \bibinfo {author} {\bibfnamefont {H.}~\bibnamefont {Kumazoe}}, \bibinfo {author} {\bibfnamefont {K.}~\bibnamefont {Nagata}}, \bibinfo {author} {\bibfnamefont {T.}~\bibnamefont {Okajima}}, \bibinfo {author} {\bibfnamefont {I.}~\bibnamefont {Akai}},\ and\ \bibinfo {author} {\bibfnamefont {M.}~\bibnamefont {Okada}},\ }\bibfield  {title} {\bibinfo {title} {Bayesian spectral deconvolution of {X}-ray absorption near edge structure discriminating between high-and low-energy domains},\ }\href@noop {} {\bibfield  {journal} {\bibinfo  {journal} {Journal of the Physical Society of Japan}\ }\textbf {\bibinfo {volume} {91}},\ \bibinfo {pages} {074009} (\bibinfo {year} {2022})}\BibitemShut {NoStop}%
\bibitem [{\citenamefont {Ueda}\ \emph {et~al.}(2023)\citenamefont {Ueda}, \citenamefont {Katakami}, \citenamefont {Yoshida}, \citenamefont {Koyama}, \citenamefont {Nakai}, \citenamefont {Mito}, \citenamefont {Mizumaki},\ and\ \citenamefont {Okada}}]{ueda2023bayesian}%
  \BibitemOpen
  \bibfield  {author} {\bibinfo {author} {\bibfnamefont {H.}~\bibnamefont {Ueda}}, \bibinfo {author} {\bibfnamefont {S.}~\bibnamefont {Katakami}}, \bibinfo {author} {\bibfnamefont {S.}~\bibnamefont {Yoshida}}, \bibinfo {author} {\bibfnamefont {T.}~\bibnamefont {Koyama}}, \bibinfo {author} {\bibfnamefont {Y.}~\bibnamefont {Nakai}}, \bibinfo {author} {\bibfnamefont {T.}~\bibnamefont {Mito}}, \bibinfo {author} {\bibfnamefont {M.}~\bibnamefont {Mizumaki}},\ and\ \bibinfo {author} {\bibfnamefont {M.}~\bibnamefont {Okada}},\ }\bibfield  {title} {\bibinfo {title} {Bayesian approach to t 1 analysis in nmr spectroscopy with applications to solid state physics},\ }\href@noop {} {\bibfield  {journal} {\bibinfo  {journal} {Journal of the Physical Society of Japan}\ }\textbf {\bibinfo {volume} {92}},\ \bibinfo {pages} {054002} (\bibinfo {year} {2023})}\BibitemShut {NoStop}%
\bibitem [{\citenamefont {Kashiwamura}\ \emph {et~al.}(2024)\citenamefont {Kashiwamura}, \citenamefont {Katakami}, \citenamefont {Yamasaki}, \citenamefont {Iwamitsu}, \citenamefont {Kumazoe}, \citenamefont {Nagata}, \citenamefont {Okajima}, \citenamefont {Akai},\ and\ \citenamefont {Okada}}]{kashiwamura2024noise}%
  \BibitemOpen
  \bibfield  {author} {\bibinfo {author} {\bibfnamefont {S.}~\bibnamefont {Kashiwamura}}, \bibinfo {author} {\bibfnamefont {S.}~\bibnamefont {Katakami}}, \bibinfo {author} {\bibfnamefont {T.}~\bibnamefont {Yamasaki}}, \bibinfo {author} {\bibfnamefont {K.}~\bibnamefont {Iwamitsu}}, \bibinfo {author} {\bibfnamefont {H.}~\bibnamefont {Kumazoe}}, \bibinfo {author} {\bibfnamefont {K.}~\bibnamefont {Nagata}}, \bibinfo {author} {\bibfnamefont {T.}~\bibnamefont {Okajima}}, \bibinfo {author} {\bibfnamefont {I.}~\bibnamefont {Akai}},\ and\ \bibinfo {author} {\bibfnamefont {M.}~\bibnamefont {Okada}},\ }\bibfield  {title} {\bibinfo {title} {Noise-robust analysis of x-ray absorption near-edge structure based on poisson distribution},\ }\href@noop {} {\bibfield  {journal} {\bibinfo  {journal} {Science and Technology of Advanced Materials: Methods}\ }\textbf {\bibinfo {volume} {4}},\ \bibinfo {pages} {2397939} (\bibinfo {year} {2024})}\BibitemShut {NoStop}%
\bibitem [{\citenamefont {Yanagita}\ and\ \citenamefont {Mikhailov}(2010)}]{yanagita2010design}%
  \BibitemOpen
  \bibfield  {author} {\bibinfo {author} {\bibfnamefont {T.}~\bibnamefont {Yanagita}}\ and\ \bibinfo {author} {\bibfnamefont {A.~S.}\ \bibnamefont {Mikhailov}},\ }\bibfield  {title} {\bibinfo {title} {Design of easily synchronizable oscillator networks using the {M}onte {C}arlo optimization method},\ }\href@noop {} {\bibfield  {journal} {\bibinfo  {journal} {Physical Review E}\ }\textbf {\bibinfo {volume} {81}},\ \bibinfo {pages} {056204} (\bibinfo {year} {2010})}\BibitemShut {NoStop}%
\bibitem [{\citenamefont {Nagata}\ and\ \citenamefont {Watanabe}(2008)}]{nagata2008asymptotic}%
  \BibitemOpen
  \bibfield  {author} {\bibinfo {author} {\bibfnamefont {K.}~\bibnamefont {Nagata}}\ and\ \bibinfo {author} {\bibfnamefont {S.}~\bibnamefont {Watanabe}},\ }\bibfield  {title} {\bibinfo {title} {Asymptotic behavior of exchange ratio in exchange {M}onte {C}arlo method},\ }\href@noop {} {\bibfield  {journal} {\bibinfo  {journal} {Neural Networks}\ }\textbf {\bibinfo {volume} {21}},\ \bibinfo {pages} {980} (\bibinfo {year} {2008})}\BibitemShut {NoStop}%
\bibitem [{\citenamefont {Iwamitsu}\ \emph {et~al.}(2021)\citenamefont {Iwamitsu}, \citenamefont {Nishi}, \citenamefont {Yamasaki}, \citenamefont {Kamezaki}, \citenamefont {Higashiyama}, \citenamefont {Yakura}, \citenamefont {Kumazoe}, \citenamefont {Aihara}, \citenamefont {Nagata}, \citenamefont {Okada} \emph {et~al.}}]{iwamitsu2021replica}%
  \BibitemOpen
  \bibfield  {author} {\bibinfo {author} {\bibfnamefont {K.}~\bibnamefont {Iwamitsu}}, \bibinfo {author} {\bibfnamefont {Y.}~\bibnamefont {Nishi}}, \bibinfo {author} {\bibfnamefont {T.}~\bibnamefont {Yamasaki}}, \bibinfo {author} {\bibfnamefont {M.}~\bibnamefont {Kamezaki}}, \bibinfo {author} {\bibfnamefont {K.}~\bibnamefont {Higashiyama}}, \bibinfo {author} {\bibfnamefont {S.}~\bibnamefont {Yakura}}, \bibinfo {author} {\bibfnamefont {H.}~\bibnamefont {Kumazoe}}, \bibinfo {author} {\bibfnamefont {S.}~\bibnamefont {Aihara}}, \bibinfo {author} {\bibfnamefont {K.}~\bibnamefont {Nagata}}, \bibinfo {author} {\bibfnamefont {M.}~\bibnamefont {Okada}}, \emph {et~al.},\ }\bibfield  {title} {\bibinfo {title} {Replica-exchange {M}onte {C}arlo method incorporating auto-tuning algorithm based on acceptance ratios for effective {B}ayesian spectroscopy},\ }\href@noop {} {\bibfield  {journal} {\bibinfo  {journal} {Journal of the Physical Society of Japan}\ }\textbf {\bibinfo {volume} {90}},\ \bibinfo {pages} {104004}
  (\bibinfo {year} {2021})}\BibitemShut {NoStop}%
\bibitem [{\citenamefont {Metropolis}\ \emph {et~al.}(1953)\citenamefont {Metropolis}, \citenamefont {Rosenbluth}, \citenamefont {Rosenbluth}, \citenamefont {Teller},\ and\ \citenamefont {Teller}}]{metropolis1953equation}%
  \BibitemOpen
  \bibfield  {author} {\bibinfo {author} {\bibfnamefont {N.}~\bibnamefont {Metropolis}}, \bibinfo {author} {\bibfnamefont {A.~W.}\ \bibnamefont {Rosenbluth}}, \bibinfo {author} {\bibfnamefont {M.~N.}\ \bibnamefont {Rosenbluth}}, \bibinfo {author} {\bibfnamefont {A.~H.}\ \bibnamefont {Teller}},\ and\ \bibinfo {author} {\bibfnamefont {E.}~\bibnamefont {Teller}},\ }\bibfield  {title} {\bibinfo {title} {Equation of state calculations by fast computing machines},\ }\href@noop {} {\bibfield  {journal} {\bibinfo  {journal} {The Journal of Chemical Physics}\ }\textbf {\bibinfo {volume} {21}},\ \bibinfo {pages} {1087} (\bibinfo {year} {1953})}\BibitemShut {NoStop}%
\bibitem [{\citenamefont {Bishop}\ and\ \citenamefont {Nasrabadi}(2006)}]{bishop2006pattern}%
  \BibitemOpen
  \bibfield  {author} {\bibinfo {author} {\bibfnamefont {C.~M.}\ \bibnamefont {Bishop}}\ and\ \bibinfo {author} {\bibfnamefont {N.~M.}\ \bibnamefont {Nasrabadi}},\ }\href@noop {} {\emph {\bibinfo {title} {Pattern recognition and machine learning}}}\ (\bibinfo  {publisher} {Springer},\ \bibinfo {year} {2006})\ p.\ \bibinfo {pages} {555}\BibitemShut {NoStop}%
\bibitem [{\citenamefont {Bernardo}\ and\ \citenamefont {Smith}(1994)}]{bernardo1994bayesian}%
  \BibitemOpen
  \bibfield  {author} {\bibinfo {author} {\bibfnamefont {J.~M.}\ \bibnamefont {Bernardo}}\ and\ \bibinfo {author} {\bibfnamefont {A.~F.}\ \bibnamefont {Smith}},\ }\href@noop {} {\emph {\bibinfo {title} {Bayesian theory}}}\ (\bibinfo  {publisher} {John Wiley \& Sons},\ \bibinfo {year} {1994})\BibitemShut {NoStop}%
\bibitem [{\citenamefont {Gelman}\ \emph {et~al.}(2013)\citenamefont {Gelman}, \citenamefont {Carlin}, \citenamefont {Stern},\ and\ \citenamefont {Rubin}}]{gelman1995bayesian}%
  \BibitemOpen
  \bibfield  {author} {\bibinfo {author} {\bibfnamefont {A.}~\bibnamefont {Gelman}}, \bibinfo {author} {\bibfnamefont {J.~B.}\ \bibnamefont {Carlin}}, \bibinfo {author} {\bibfnamefont {H.~S.}\ \bibnamefont {Stern}},\ and\ \bibinfo {author} {\bibfnamefont {D.~B.}\ \bibnamefont {Rubin}},\ }\href@noop {} {\emph {\bibinfo {title} {Bayesian data analysis}}},\ \bibinfo {edition} {3rd}\ ed.\ (\bibinfo  {publisher} {Chapman and Hall/CRC},\ \bibinfo {year} {2013})\BibitemShut {NoStop}%
\bibitem [{\citenamefont {Wenden}\ \emph {et~al.}(2012)\citenamefont {Wenden}, \citenamefont {Toner}, \citenamefont {Hodge}, \citenamefont {Grima},\ and\ \citenamefont {Millar}}]{wenden2012spontaneous}%
  \BibitemOpen
  \bibfield  {author} {\bibinfo {author} {\bibfnamefont {B.}~\bibnamefont {Wenden}}, \bibinfo {author} {\bibfnamefont {D.~L.}\ \bibnamefont {Toner}}, \bibinfo {author} {\bibfnamefont {S.~K.}\ \bibnamefont {Hodge}}, \bibinfo {author} {\bibfnamefont {R.}~\bibnamefont {Grima}},\ and\ \bibinfo {author} {\bibfnamefont {A.~J.}\ \bibnamefont {Millar}},\ }\bibfield  {title} {\bibinfo {title} {Spontaneous spatiotemporal waves of gene expression from biological clocks in the leaf},\ }\href@noop {} {\bibfield  {journal} {\bibinfo  {journal} {Proceedings of the National Academy of Sciences}\ }\textbf {\bibinfo {volume} {109}},\ \bibinfo {pages} {6757} (\bibinfo {year} {2012})}\BibitemShut {NoStop}%
\bibitem [{\citenamefont {Muranaka}\ and\ \citenamefont {Oyama}(2016)}]{muranaka2016heterogeneity}%
  \BibitemOpen
  \bibfield  {author} {\bibinfo {author} {\bibfnamefont {T.}~\bibnamefont {Muranaka}}\ and\ \bibinfo {author} {\bibfnamefont {T.}~\bibnamefont {Oyama}},\ }\bibfield  {title} {\bibinfo {title} {Heterogeneity of cellular circadian clocks in intact plants and its correction under light-dark cycles},\ }\href@noop {} {\bibfield  {journal} {\bibinfo  {journal} {Science Advances}\ }\textbf {\bibinfo {volume} {2}},\ \bibinfo {pages} {e1600500} (\bibinfo {year} {2016})}\BibitemShut {NoStop}%
\bibitem [{\citenamefont {Watanabe}\ \emph {et~al.}(2023)\citenamefont {Watanabe}, \citenamefont {Muranaka}, \citenamefont {Nakamura}, \citenamefont {Isoda}, \citenamefont {Horikawa}, \citenamefont {Aiso}, \citenamefont {Ito},\ and\ \citenamefont {Oyama}}]{watanabe2023non}%
  \BibitemOpen
  \bibfield  {author} {\bibinfo {author} {\bibfnamefont {E.}~\bibnamefont {Watanabe}}, \bibinfo {author} {\bibfnamefont {T.}~\bibnamefont {Muranaka}}, \bibinfo {author} {\bibfnamefont {S.}~\bibnamefont {Nakamura}}, \bibinfo {author} {\bibfnamefont {M.}~\bibnamefont {Isoda}}, \bibinfo {author} {\bibfnamefont {Y.}~\bibnamefont {Horikawa}}, \bibinfo {author} {\bibfnamefont {T.}~\bibnamefont {Aiso}}, \bibinfo {author} {\bibfnamefont {S.}~\bibnamefont {Ito}},\ and\ \bibinfo {author} {\bibfnamefont {T.}~\bibnamefont {Oyama}},\ }\bibfield  {title} {\bibinfo {title} {A non-cell-autonomous circadian rhythm of bioluminescence reporter activities in individual duckweed cells},\ }\href@noop {} {\bibfield  {journal} {\bibinfo  {journal} {Plant Physiology}\ }\textbf {\bibinfo {volume} {193}},\ \bibinfo {pages} {677} (\bibinfo {year} {2023})}\BibitemShut {NoStop}%
\bibitem [{\citenamefont {Gardiner}(2009)}]{gardiner2009stochastic}%
  \BibitemOpen
  \bibfield  {author} {\bibinfo {author} {\bibfnamefont {C.}~\bibnamefont {Gardiner}},\ }\href@noop {} {\emph {\bibinfo {title} {Stochastic methods}}},\ \bibinfo {edition} {4th}\ ed.\ (\bibinfo  {publisher} {Springer Berlin Heidelberg},\ \bibinfo {year} {2009})\BibitemShut {NoStop}%
\bibitem [{\citenamefont {Yue}\ and\ \citenamefont {Gottwald}(2024)}]{yue2024stochastic}%
  \BibitemOpen
  \bibfield  {author} {\bibinfo {author} {\bibfnamefont {W.}~\bibnamefont {Yue}}\ and\ \bibinfo {author} {\bibfnamefont {G.~A.}\ \bibnamefont {Gottwald}},\ }\bibfield  {title} {\bibinfo {title} {A stochastic approximation for the finite-size {K}uramoto--{S}akaguchi model},\ }\href@noop {} {\bibfield  {journal} {\bibinfo  {journal} {Physica D: Nonlinear Phenomena}\ }\textbf {\bibinfo {volume} {468}},\ \bibinfo {pages} {134292} (\bibinfo {year} {2024})}\BibitemShut {NoStop}%
\bibitem [{\citenamefont {Sakaguchi}\ and\ \citenamefont {Kuramoto}(1986)}]{sakaguchi1986soluble}%
  \BibitemOpen
  \bibfield  {author} {\bibinfo {author} {\bibfnamefont {H.}~\bibnamefont {Sakaguchi}}\ and\ \bibinfo {author} {\bibfnamefont {Y.}~\bibnamefont {Kuramoto}},\ }\bibfield  {title} {\bibinfo {title} {A soluble active rotater model showing phase transitions via mutual entertainment},\ }\href@noop {} {\bibfield  {journal} {\bibinfo  {journal} {Progress of Theoretical Physics}\ }\textbf {\bibinfo {volume} {76}},\ \bibinfo {pages} {576} (\bibinfo {year} {1986})}\BibitemShut {NoStop}%
\bibitem [{\citenamefont {Snyder}\ \emph {et~al.}(2021)\citenamefont {Snyder}, \citenamefont {Callaham}, \citenamefont {Brunton},\ and\ \citenamefont {Kutz}}]{snyder2021data}%
  \BibitemOpen
  \bibfield  {author} {\bibinfo {author} {\bibfnamefont {J.}~\bibnamefont {Snyder}}, \bibinfo {author} {\bibfnamefont {J.~L.}\ \bibnamefont {Callaham}}, \bibinfo {author} {\bibfnamefont {S.~L.}\ \bibnamefont {Brunton}},\ and\ \bibinfo {author} {\bibfnamefont {J.~N.}\ \bibnamefont {Kutz}},\ }\bibfield  {title} {\bibinfo {title} {Data-driven stochastic modeling of coarse-grained dynamics with finite-size effects using {L}angevin regression},\ }\href@noop {} {\bibfield  {journal} {\bibinfo  {journal} {Physica D: Nonlinear Phenomena}\ }\textbf {\bibinfo {volume} {427}},\ \bibinfo {pages} {133004} (\bibinfo {year} {2021})}\BibitemShut {NoStop}%
\end{thebibliography}%

\end{document}